\documentclass[aps,prb,reprint,superscriptaddress]{revtex4-2}
\usepackage{amsmath}
\usepackage{amssymb}
\usepackage{graphicx}
\usepackage{url}
\usepackage[usenames,dvipsnames]{color}

\newcommand{\degrees}{\mbox{$^\circ$}}
\newcommand{\celcius}{\mbox{\,\degrees{C}}}

\newcommand{\cm}{\mbox{\,cm$^{-1}$}}
\newcommand{\ev}{\mbox{{\,eV}}}
\newcommand{\mev}{\mbox{\,meV}}

\newcommand{\bc}{\begin{center}}
\newcommand{\ec}{\end{center}}

\newcommand{\sint}{\mbox{${\rm S}_i$}}
\newcommand{\vs}{\mbox{${\rm V_S}$}}
\newcommand{\vmo}{\mbox{${\rm V_{Mo}}$}}
\newcommand{\cint}{\mbox{${\rm C}_i$}}
\newcommand{\cintAboveMo}{\mbox{${\rm C}_i^{{\rm Mo}}$}}
\newcommand{\cintMoC}{\mbox{${\rm C}_i^{{\rm MoC}}$}}
\newcommand{\cintAboveS}{\mbox{${\rm C}_i^{{\rm S}}$}}
\newcommand{\cintAboveVoid}{\mbox{${\rm C}_i^{{\rm v}}$}}
\newcommand{\cintBridge}{\mbox{${\rm C}_i^{\rm Bridge}$}}
\newcommand{\cs}{\mbox{$\rm C_S$}}
\newcommand{\cmo}{\mbox{$\rm C_{Mo}$}}
\newcommand{\cmothreefold}{\mbox{${\rm C}_{\rm Mo}^{\rm 3-fold}$}}
\newcommand{\cmofourfold}{\mbox{${\rm C}_{\rm Mo}^{\rm 4-fold}$}}
\newcommand{\cmosixfold}{\mbox{${\rm C}_{\rm Mo}^{\rm 6-fold}$}}
\newcommand{\ccmo}{\mbox{$\rm (C_2)_{Mo}$}}
\newcommand{\cssi}{\mbox{\cs-\sint}}

\newcommand{\mos}{\mbox{MoS$_2$}}

\begin{document}

%\begin{frontmatter}

\title{Reevaluating the electrical impact of atomic carbon impurities in {\mos}}

%\affiliation[inst1]{organization={School of Mathematics, Statistics and Physics, Newcastle University},
 %city={Newcastle upon Tyne},
 %postcode={NE1 7RU}, 
 %country={UK}}
%\affiliation[inst2]{organization={College of Science and Arts, Najran University}, city={Najran}, 
%country={Saudi Arabia}}
%\author[inst1]{James Ramsey}
%\author[inst1,inst2]{Faiza Alhamed}
%\author[inst1]{J. P. Goss}
%\author[inst1]{P. R. Briddon}
%\author[inst1]{M. J. Rayson}

\author{James Ramsey}
\affiliation{School of Mathematics, Statistics and Physics, Newcastle University, Newcastle upon Tyne, NE1 7RU, UK}

\author{Faiza Alhamed}
\affiliation{School of Mathematics, Statistics and Physics, Newcastle University, Newcastle upon Tyne, NE1 7RU, UK}
\affiliation{College of Science and Arts, Najran University, Najran, Saudi Arabia}

\author{J. P. Goss}
\affiliation{School of Mathematics, Statistics and Physics, Newcastle University, Newcastle upon Tyne, NE1 7RU, UK}
 
\author{P. R. Briddon}
\affiliation{School of Mathematics, Statistics and Physics, Newcastle University, Newcastle upon Tyne, NE1 7RU, UK}
 
\author{M. J. Rayson}
\affiliation{School of Mathematics, Statistics and Physics, Newcastle University, Newcastle upon Tyne, NE1 7RU, UK}

\begin{abstract}
Transition metal dichalcogenides, a family of two-dimensional compounds, are of interest for a range of technological applications.  {\mos}, the most researched member of this family, is hexagonal, from which monolayers may be isolated. 
Under ambient conditions and during growth/processing, contamination by impurities can occur, of which carbon is significant due to its presence in the common growth techniques. 
We have performed extensive computational investigations of carbon point defects, examining substitutional and interstitial locations. 
Previously unreported thermodynamically stable configurations, four-fold co-ordinated mono-carbon and di-carbon substitutions of Mo, and a complex of carbon substitution of sulfur bound to interstitial sulfur have been identified.
We find no evidence to support recent assertions that carbon defects are responsible for electrical doping of {\mos}, finding all energetically favorable forms have only deep charge transition levels and would act as carrier traps.
To aid unambiguous identification of carbon defects, we present electronic and vibrational data for comparison with spectroscopy.
\end{abstract}

%\includegraphics[width=0.48\textwidth]{StructureImages/C_Mo_4fold_cutout.png}

%\item Something

%\begin{keyword}
%TMD \sep point defect \sep modeling
%%\PACS 0000 \sep 1111
%%\MSC 0000 \sep 1111
%\end{keyword}

%\end{frontmatter}

\maketitle

\section{Introduction \label{sec:intro}}

Since the experimental isolation of monolayer graphene\,\cite{novoselov-S-306-666}, 2D van der Waals-layered materials have been a promising area of research interest due to their electrical and mechanical properties. 
Combining this with their atomic length scales, they are positioned as strong candidates to overcome the scaling limits\,\cite{frank-PIEEE-259} of magnetic data storage devices\,\cite{cong-SR-5-9361}, photovoltaic cells\,\cite{li-CSR-47-4981} and conventional silicon-based electronics\,\cite{yin-NPJ2DMA-2397,radisavljevic-ACSN-9934,jariwala-ACSN-8-1102}. 
The zero bandgap limits graphene's application, whereas semiconducting transition metal dichalcogenides (TMDs) exhibit suitable bandgaps in the range 1.1--2.0\ev\,\cite{ryder-ACSN-10-3900}, tunable by the number of layers; the {\mos} bandgap varies monotonically with thickness between 1.23{\ev} (indirect)\,\cite{kam-JPCY-86-463} in bulk, to 1.9{\ev}\,\cite{gusakova-PSSA-214, mak-PRL-105, island-NS-8-2589, huang-NCOMM-6} for the monolayer (direct). 
%\cite{camacho-JC-234-182,jariwala-ACSN-8-1102,he-NN-10-497,koperski-NN-10-503,koperski-NN-10-503,li-NN-9-372, barthelmi-APL-117}
{\mos} also benefits from a room-temperature carrier mobility\,\cite{radisavljevic-NN-6-147} $>$200\,{cm$^2$/V.s}, comparable to Si, an in-plane Young's modulus exceeding that of steel\,\cite{castellanos-AM-24-772} and a solar photon absorption an order of magnitude greater than Si or GaAs\,\cite{bernardi-NL-13-3664} due to its high refractive index\,\cite{jariwala-ACSP-12-2962}.

Presently, the performance of TMD-based optoelectronic devices does not meet theoretical predictions. 
For example, photovoltaic cells deliver power conversion efficiencies lower than their theoretical maximum\,\cite{jariwala-ACSP-12-2962} by factors of 5 or greater, even for state-of-the-art devices\,\cite{kim-ACSN-16-8827, tsai-AM-29, kwak-ADVMI-5}. 
The main reasons for under-performance include device-level factors such as Fermi-level pinning\,\cite{nassiri-NCOMM-12}, contamination or corrugation at TMD-TMD interfaces\,\cite{tsai-AM-29} and inefficient exciton separation\,\cite{kim-ACSN-16-8827} due to charge-carrier trapping at point defects\,\cite{tongay-SR-3, nassiri-NL-21-3443}. 
Characterizing contamination and point defects is therefore crucial.

Native point defects are dominant over impurities in {\mos}. In Mo-rich (S-lean) growth conditions, literature data\,\cite{zhou-NL-13-2615,xu-MTC-22-100772,tan-PRM-4-064004,haldar-PRB-92-235408,komsa-PRB-91-125304,noh-PRB-89-205417} consistently indicate the sulfur vacancy,\,{\vs}, is pervasive due to its energetically favorable formation energy, $E_f$, at 1.5{\ev}\,\cite{zhou-NL-13-2615,xu-MTC-22-100772}.
in combination with its sizable ($\gtrapprox 2${\ev}) diffusion barrier\,\cite{gusakov-PSSB-259-2100479} for all stable charge states\,\cite{komsa-PRB-91-125304}, rendering it immobile under ambient, processing and growth conditions. 
In the Mo-lean (S-rich) condition, the sulfur interstitial,\,{\sint}, presents as another stable center with low $E_f$ (0.9{\ev}\,\cite{santosh-NA-375703,noh-PRB-89-205417}), which is immobile at room temperature due to its appreciable diffusion barrier (1.7{\ev}\,\cite{komsa-PRB-91-125304,gusakov-PSSB-259-2100479}).

As for impurities, it has been suggested that carbon is unintentionally substituted into the material from low purity Mo precursors during chemical vapor deposition (CVD)\,\cite{liang-ACSAEM-2-1055}.
It is also proposed that carbon is co-deposited during metal-organic CVD\,\cite{schaefer-CM-4474} from pyrolysis of the metal organic precursors\,\cite{ghiami-SUR-6-351}, followed by subsequent substitution into the host material under atmospheric conditions\,\cite{ghiami-SUR-6-351}. 
Carbon is therefore a notable contaminant to understand the properties of as-grown {\mos}, particularly for scalable applications.
% I think the following pair of sentences are no longer required. A coherent and compelling narrative has been drawn without it. The value in our paper is no longer simply providing a more comprehensive overview of carbon defects in MoS2, but also dispelling spurious claims of electrical conductivity from the literature. Hence, I've commented this out for now and pushed the focus onto the overarching conductivity narrative
%First-principles modeling of carbon-containing point defects in {\mos} falls broadly into three categories: studies which cover a wide variety of elemental defect-species, of which carbon is included\,\cite{li-AIPA-5,ataca-JPCC-115-13303,he-APL-96}; studies of molecular carbon defects\,\cite{miao-SAI-27-101580,enujekwu-ASS-542-148556,dacunha-PCCP-21-11168}; and experimental investigations where the focus is upon applied, functionalized TMDs, using first-principle calculations to supplement/corroborate the results\,\cite{park-NPJ2DMA-7-60,xiao-STE-858-159587,ma-JCIS-569-89}. This has resulted in limited investigation of both substitutional and interstitial carbon point defects, which would be remedied by a dedicated first-principles investigation, and it is notably absent from the literature at present.
%%%%%%%%%%%%%%%%%%%%%%%%%%%%%%%%%%%%%

Carbon is also considered a prime candidate for a source of n-type conductivity\,\cite{schaefer-CM-4474,park-NPJ2DMA-7-60,radisavljevic-NN-6-147,kim-NCOMM-1038,yin-ACSN-74,schmidt-NL-1909,li-SML-63,lee-AM-2320,liu-NL-1538,baugher-NL-4212,ahn-SR-7} in field-effect transistors constructed of as-grown {\mos}.
However, the precise origin of the conductivity is disputed. 
Different authors suggest Cl or Br substitution for S\,\cite{yin-ACSN-74}, Re substitution for Mo\,\cite{komsa-PRB-91-125304}, hydrogenated native defects\,\cite{singh-PRB-121201}, and carbon\,\cite{schaefer-CM-4474,park-NPJ2DMA-7-60}, specifically interstitial impurities\,\cite{park-NPJ2DMA-7-60}, could be potential causes. %of the behaviour. 
In particular, Ref.\,\cite{park-NPJ2DMA-7-60} attributed n-type conductivity to carbon by experimental observation of a concurrent p-to-n-type transition with increasing carbon deposition time. 
This is identified by a relative shift of the Fermi level away from the valence band edge in angle-resolved photon-emission spectroscopy. 
A similar dependence of nanometre-scale defect concentration (notably not atomic scale) on carbon deposition time is also identified. 
These observations were sought to be explained using first-principles calculations, noting that some carbon interstitial structures shift the Fermi energy towards the conduction band relative to the pure material\,\cite{park-NPJ2DMA-7-60}.
In the candidate structures, an additional carbon atom is located above a Mo-site, approximately in-plane with three sulfur neighbors, forming four-fold coordination, ({\cintAboveMo}, Fig.\,\ref{fig:geometry_structure_of_C_defects}(f), Refs.\,\cite{park-NPJ2DMA-7-60,auni-ACSANM-28098,he-APL-96,li-AIPA-5,aghajanian-ARXIV-1805}), or alternatively with C bridging between a Mo and a S atom in a two-fold coordination, ({\cintBridge}, Fig.\,\ref{fig:geometry_structure_of_C_defects}(g)).
However, such an assignment is rendered problematic when noting that these structures are not the equilibrium structure\,\cite{ataca-JPCC-115-13303}, 
with C lying in the Mo-layer coordinated to three Mo ({\cintMoC}, Fig.\,\ref{fig:geometry_structure_of_C_defects}(h), Ref\,\cite{ataca-JPCC-115-13303}) being more energetically stable.
% For the study\,\cite{ataca-JPCC-115-13303} that included {\cintMoC}, this is determined to be the most energetically favorable form, followed by {\cintAboveMo} which is close in energy to {\cintBridge},\cite{park-NPJ2DMA-7-60}.
In fact, several more recent publications\,\cite{auni-ACSANM-28098,li-AIPA-5,liang-ACSAEM-2-1055,aghajanian-ARXIV-1805} omit this low-energy configuration from their analysis.
%Furthermore, there is a large spread amongst these adsorption energy data ($>2{\ev}$ for a given geometry) with little explanation given, likely due to a largely incomplete description of the utilized reference energies.
This highlights a need to establish more definitively the equilibrium behavior of carbon interstitials, and their properties. 

Ultimately, the impact an impurity has upon electrical conductivity must correspond to a mechanism for the release or capture of free carriers.
If carbon has a direct role in electrical doping, this should be both evident from the band structure, and relate to thermodynamically favorable arrangements.

In this paper, we present the results of a detailed investigation of carbon point-defects in monolayer {\mos} using density functional theory (DFT). 
We report geometries, energetic stabilities, electronic structures, magnetic states and vibrational modes, providing routes to experimental observation. 
Our focus is upon a robust evaluation of relative energetics of different forms of carbon contamination, and to challenge the conclusions that simple centers, especially interstitial carbon, can be responsible for electrical doping.
The inclusion of spectroscopically observable properties, both from the band structure and from the vibrational modes, is with the aim of providing direct routes to experimental identification of the microstructures of carbon defects in {\mos}.

\section{Computational method \label{sec:method}}

Calculations were performed using the AIMPRO\,\cite{rayson-PRB-80-205104,jones-SSM-51-287} DFT modeling software package, within the PBE generalized gradient approximation\,\cite{perdew-PRL-77-3865}, and the Grimme-D2 van der Waals (vdW) correction\,\cite{grimme-JCC-1787}.
Pseudopotentials\,\cite{hartwigsen-PRB-58-3641,krack-TCA-114-145} were utilized with 4, 14 and 6 electrons in the valence sets for C, Mo and S, respectively, eliminating core electrons.
Kohn-Sham functions were expanded using atom-centered, independent Gaussian orbital functions\,\cite{goss-TAP-104-69}, using $s$-, $p$- and $d$-type functions with four widths amounting to 40 functions per atom.
Hamiltonian matrix-elements were generated using a plane-wave expansion\,\cite{rayson-PRB-80-205104} of the charge-density and Kohn-Sham potential with a minimum of 150\,Ha cut-off, yielding well converged total energies (better than 0.1\,meV/atom) with respect to this parameter. 

For all calculations except those for minimum energy paths (see below), defects were introduced to a base-supercell comprised of 300-atoms made up from $10\times10$ primitive unit cells.
Integration over the Brillouin-zone is by Monkhorst-Pack sampling\,\cite{monkhorst-PRB-13-5188}; the primitive pristine monolayer sampling is $15\times15$, with all non-primitive cells modeled with equal or greater sampling densities.
To determine an estimate of the impact of the choice of sampling scheme, the total energy of pristine {\mos} was determined using a range of sampling densities.
We find that using the $15\times15$ mesh results absolute total energies are converged to $\ll$\,1{\mev}/atom.

Equilibrium structures are obtained by optimizing geometries until the maximum final force is less than 1\,mHa/bohr.
Using this approach, the in-plane lattice constant of pristine monolayer {\mos} is found to be 3.20\,{\AA}, in agreement with previous reports\,\cite{noh-PRB-89-205417,haldar-PRB-92-235408}.
The lattice constant perpendicular to {\mos} is 15{\,\AA} to minimize interactions between periodic images.
The Mo-S bond lengths and electronic bandgap of monolayer {\mos} were found to be 2.42\,{\AA} and 1.64\,eV, respectively, also in agreement with previous calculations (2.4\,{\AA}\,\cite{le-RSCA-2046,ataca2011mechanical} and 1.7\,eV\,\cite{singh-PRB-121201,hieu2018first}).

Formation energies\,\cite{zhang-PRL-67-2339} of system $X$, $E_f(X)$, are obtained using $ E_f(X)=E_t(X)-\sum_in_i\mu_i$, where $E_t(X)$ is the total energy and $n_i$ is the number of species $i$ with chemical potential $\mu_i$ in the pristine system of identical supercell dimension.
Taking the equilibrium phase as pure monolayer {\mos} which must be stable relative to their elemental components, $\mu_{\rm Mo}=E_t\left({\rm MoS_2}\right) - 2\mu_{\rm S}$, with $\mu_{\rm mo}$ bounded by the energy per atom of body-centered cubic Mo (Mo-rich limit) and $\mu_{\rm S}$ the energy per atom of bulk sulfur (Mo-lean limit).
We define the relative chemical potential of Mo as $\Delta\mu_{\rm Mo}=0$ for Mo-lean conditions, and the carbon atomic chemical potential is taken to be the energy per atom of pure diamond.

Carbon adsorption energies, $E_{\rm ad}$, are formation energies obtained with $\mu_{\rm C}$ being the energy of a free carbon atom, $E_{\rm C}$. 
The ground-state energy of a carbon atom is challenging because the many-body wavefunctions are not well represented by the single determinental formulation in DFT.  
To accommodate this issue, the experimental cohesive energy of diamond ($E_{\rm coh}=-7.43${\ev}/atom\,\cite{goto-JPSJ-895,gaydon-PRSLA-374}) is used to infer the atomic energy, i.e. $E_{\rm C}=E_{\rm diamond} - E_{\rm coh}$, where $E_{\rm diamond}$ is the calculated energy per atom of diamond.
Based upon this estimate of the carbon energy, our calculated values obtained by fractional filling of the $2p$-states for a single atom represent overestimates by 0.9 and 2.5{\ev} for spin-polarized and spin-averaged configurations, respectively.
Hence, how the atom energy is obtained will have a significant impact upon estimated of adsorption energies.

Minimum energy paths were calculated using the climbing nudged-elastic-band method\,\cite{henkelman-JCP-113-9901,henkelman-JCP-113-9978} utilizing supercells of $4\times4$ or $6\times6$ primitive cells and a minimum of 9 images. The saddle-points were allowed to climb after the third iteration. The larger cell is used for dissociation of a sulfur interstitial from a complex with {\cs} to allow for the separation of the component parts. Convergence tests on selected reactions indicate activation energies determined in the smaller cell are uncertain to 10s of meV.

Finally, local vibrational modes were modeled within the harmonic approximation, with dynamical-matrix elements found using a finite difference approximation for the second-derivatives of the energy with respect to displacement\,\cite{jones-PRB-50-8378}.

\section{Results}

We have examined carbon substitutions at the Mo or S sites and as an intersticy. 
To provide for reaction energies, we have also modeled mono-vacancies and sulfur-interstitial defects.
We report here geometric, energetic, electronic and vibrational data, starting with the defect geometries.

\subsection{Defect Geometries}\label{Section:DefectGeometries}

\begin{figure*}[!htb]
\begin{minipage}{0.32\textwidth}(a) {\cmothreefold}\\\includegraphics[width=\textwidth]{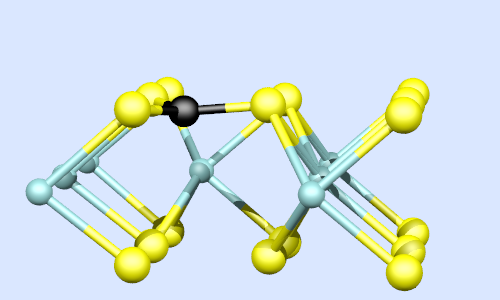}\end{minipage}\hfill
\begin{minipage}{0.32\textwidth}(b) {\cmo}\\\includegraphics[width=\textwidth]{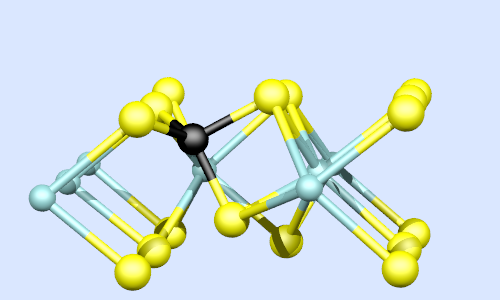}\end{minipage}\hfill
\begin{minipage}{0.32\textwidth}(c) {\ccmo}\\\includegraphics[width=\textwidth]{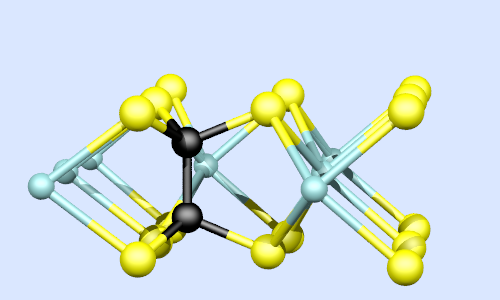}\end{minipage}\\
\begin{minipage}{0.32\textwidth}(d) {\cs}\\\includegraphics[width=\textwidth]{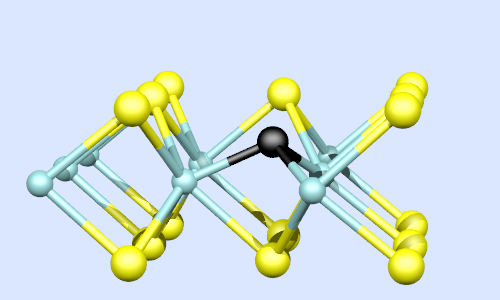}\end{minipage}\hfill
\begin{minipage}{0.32\textwidth}(e) {\cintAboveS}\\\includegraphics[width=\textwidth]{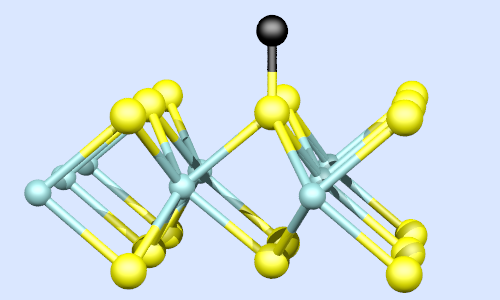}\end{minipage}\hfill
\begin{minipage}{0.32\textwidth}(f) {\cintAboveMo}\\\includegraphics[width=\textwidth]{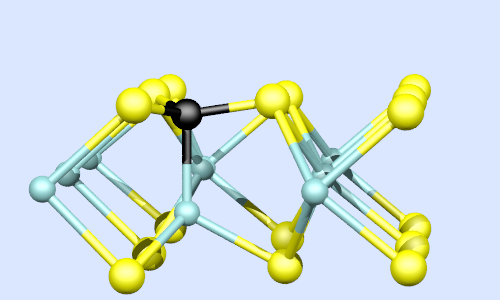}\end{minipage}\\
\begin{minipage}{0.32\textwidth}(g) {\cintBridge}\\\includegraphics[width=\textwidth]{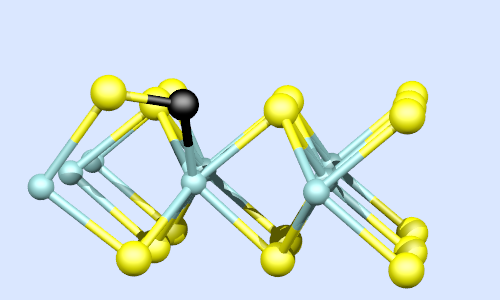}\end{minipage}\hfill
\begin{minipage}{0.32\textwidth}(h) {\cintMoC}\\\includegraphics[width=\textwidth]{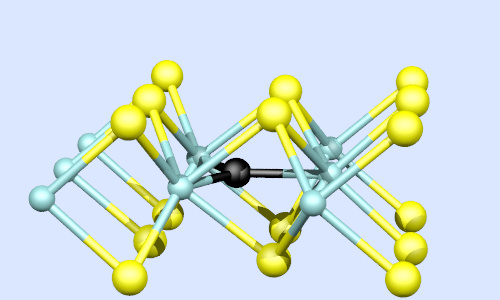}\end{minipage}\hfill
\begin{minipage}{0.32\textwidth}(i) ({\cssi})$^{\rm HS}$\\\includegraphics[width=\textwidth]{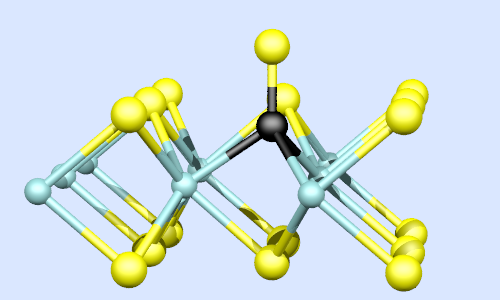}\end{minipage}\\
\begin{minipage}{0.32\textwidth}(j) ({\cssi})$^{\rm LS}$\\\includegraphics[width=\textwidth]{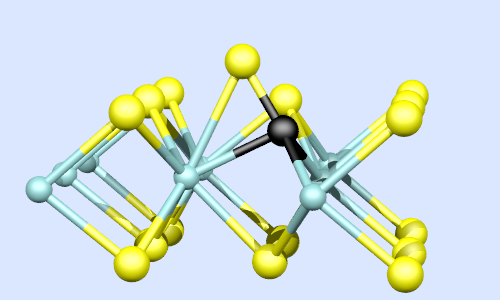}\end{minipage}\hfill
\begin{minipage}{0.32\textwidth}(k) Pristine {\mos}\\\includegraphics[width=\textwidth]{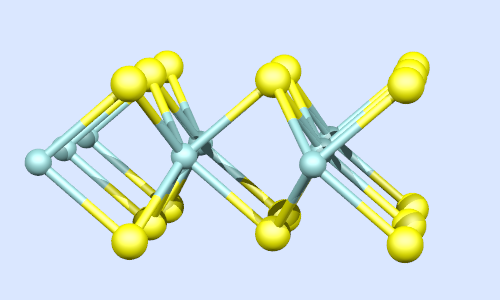}\end{minipage}\hfill
\begin{minipage}{0.32\textwidth}(l) Pristine {\mos} plan-view\\\includegraphics[width=\textwidth]{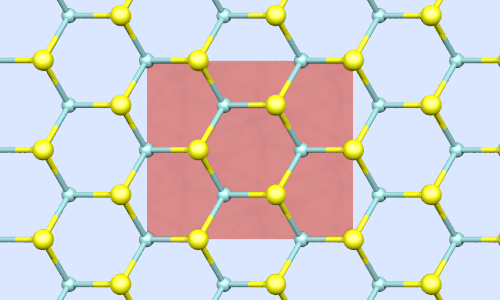}\end{minipage}\\
 \caption{Schematic structures of carbon defects in monolayer {\mos}. (a) and (b) show the three- and four-fold co-ordinated configurations of {\cmo}, (c) the di-carbon substitution, {\ccmo} (d) {\cs}, and (e)--(h) show C$_{i}$ in four configurations where the carbon is above S, above Mo, bridging between the Mo and S sites, and in the Mo-plane, respectively. (i) and (j) illustrate {\cssi} in high (HS) and low symmetry (LS) configurations. Blue, yellow and black spheres represent Mo, S and C, respectively. (k) illustrates pristine {\mos} for comparison and (l) shows in plan view the depicted material section (shaded area) for each of the structures. Depictions of the centers from plan-view are available in the supplementary information.
 }
\label{fig:geometry_structure_of_C_defects}
\end{figure*}

\begin{table*}[!hbt]
 \centering
 \caption{Symmetry, co-ordination and selected geometric parameters of C-containing point defects in monolayer {\mos}. Distances are in {\AA} and angles are quoted to the nearest degree. For comparison, for pristine {\mos} the calculated Mo--S distance is $2.42${\,\AA} and $\angle$Mo-S-Mo is 83\degrees{}.}
 \label{tab:geometry}
 \begin{tabular}{ccccrlrl}
 %\toprule
 Defect & Sym{.} & Fig.\,\ref{fig:geometry_structure_of_C_defects}& C co-ord{.} & \multicolumn{2}{c}{Geometric Quantities} \\ \hline
{\cmo}& $C_{3v}$ & (a) & 3 & $\begin{array}{r}\text{C--S}\\\angle\text{SCS}\\\end{array}$ & $\begin{array}{l}1.74\\119\end{array}$\\
{\cmo} & $C_{1h}$ & (b) & 4 & $\begin{array}{r}\text{C--S}\\\angle\text{SCS}\\\end{array}$ & $\begin{array}{l}1.85\\ 93, 108, 109\times2, 118\times2\\ \end{array}$\\
{\ccmo} & $D_{3h}$ & (c) & 4 & $\begin{array}{r}\text{C--S, C--C}\\ \angle\text{SCS}, \angle\text{CCS}\\\end{array}$& $\begin{array}{l}1.83, 1.57\\ 106, 113\\ \end{array}$ &&\\
{\cs}& $C_{3v}$ & (d) & 3 & $\begin{array}{r}\text{C--Mo}\\ \angle\text{MoCMo}\\ \end{array} $ & $\begin{array}{l}2.02\\ 104\\ \end{array}$ \\
{\cintAboveS} & $C_{3v}$ & (e) & 1 & $\begin{array}{r}\text{C--S}\\ \angle\text{SCMo}, \angle\text{MoCMo}\\ \end{array}$& $\begin{array}{l}1.68\\ 130, 84\\ \end{array}$\\
{\cintAboveMo} & $C_{3v}$ & (f) & 4 & $\begin{array}{r}\text{C--S, C--Mo}\\\angle\text{CSMo}, \angle\text{MoCMo}\\ \end{array}$& $\begin{array}{l}1.81, 2.06\\ 118, 99\\ \end{array}$ \\
{\cintBridge} & $C_{1}$ & (g) & 2 & $\begin{array}{r}\text{C--S, C--Mo}\\\angle\text{SCMo}\\ \end{array}$& $\begin{array}{l}1.64, 2.14\\116 \\ \end{array}$ \\
{\cintMoC} & $D_{3h}$ & (h) & 3 & $\begin{array}{r}\text{C--Mo}\\\angle\text{MoCMo}\\ \end{array}$& $\begin{array}{l}2.03\\ 120\\ \end{array}$\\
({\cssi})$^{\rm HS}$& $C_{3v}$ & (i) & 3 & $\begin{array}{r}\text{C--Mo, C--S}\\ \angle\text{MoCMo},\angle\text{SCMo}\\ \end{array}$ & $\begin{array}{l}2.20, 1.64\\91, 124 \\ \end{array}$ \\
({\cssi})$^{\rm LS}$& $C_{1h}$ & (j) & 4 & $\begin{array}{r}\text{C--Mo, C--S}\\\angle\text{MoCMo}, \angle\text{MoCS}, \angle\text{CSMo} \\ \end{array}$& $\begin{array}{l} (2.12\times 2, 2.23), 1.66 \\ (93\times 2, 97), (132\times 2, 86), 56\\ \end{array}$ \\
 \end{tabular}
\end{table*}

Fig.\,\ref{fig:geometry_structure_of_C_defects} illustrates carbon defects with associated structural parameters listed in Table~\ref{tab:geometry}. Pristine {\mos} is illustrated in Fig.\,\ref{fig:geometry_structure_of_C_defects}(k-l) for comparison with each impurity.

Substitution of Mo by C, {\cmo}, directly onto the Mo site results in a $D_{3h}$ point-group symmetry and six-fold effective co-ordination of C (\cmosixfold). 
The optimized Mo--S distance is 2.29{\,\AA}, and we note the C--Mo distances of 3.23{\,\AA} is much greater than the 2.18{\,\AA} bond-length in $\gamma$-MoC\,\cite{kuo-N-170-245}. 
However, {\cmosixfold} is not the equilibrium structure, but rather a local energy maximum.
Previous studies\,\cite{liang-ACSAEM-2-1055} suggest {\cmo} is three-fold-co-ordinated (\cmothreefold, (Fig.\,\ref{fig:geometry_structure_of_C_defects}(a)), arising from C-displacement from the Mo-site along [0001], rendering three S-sites under-co-ordinated.
{\cmothreefold} is 2.6{\ev} lower in energy than  {\cmosixfold}, with C--S bond-lengths of 1.74{\,\AA}, much closer to typical C--S bond-lengths in organic compounds.
Both {\cmosixfold} and {\cmothreefold} energetically favor a spin-triplet electronic configuration.

Our findings are in line with the literature\,\cite{liang-ACSAEM-2-1055}, but we find a $C_{1h}$-symmetry four-fold co-ordinated arrangement 0.6{\ev} lower still in energy (\cmofourfold, Fig.\,\ref{fig:geometry_structure_of_C_defects}(b)).
Considering the six S-neighbors of {\cmosixfold} to be in two groups of three on either side of the Mo-plane, the structure of {\cmofourfold} can be understood as carbon forming bonds with one set of three (as in {\cmothreefold}), with the fourth C--S bond forming  with one S-atom in the other group.
Although not constrained by its $C_{1h}$ symmetry, we find the four C--S bond-lengths (Table~\ref{tab:geometry}) are the same length to within 0.5\,pm, suggesting that all four bonds are of the same type.

Both {\cmothreefold} and {\cmofourfold} contain under-co-ordinated S-sites, suggesting that placing C in both S-layers may also be favorable.
We have modeled this di-carbon center, {\ccmo}, the optimized structure being shown in Fig.\,\ref{fig:geometry_structure_of_C_defects}(c).
Both carbon atoms are four-fold co-ordinated and the equilibrium structure retains a high symmetry ($D_{3h}$).
The C--C bond is 1.57{\,\AA}, consistent with a $sp^3$ single bond configuration and we note the C--S bonds are similar in length to the four-fold co-ordinated {\cmo} arrangement, indicating a comparable bonding.

Substitution for the sulfur site, {\cs} (Fig.\,\ref{fig:geometry_structure_of_C_defects}(d)), also results in a high-symmetry configuration, with the three-fold co-ordination of carbon. 
Our findings are in agreement with the literature\,\cite{park-NPJ2DMA-7-60,liang-ACSAEM-2-1055,auni-ACSANM-28098}. 
The C--Mo bond lengths are 2.02{\,\AA}, much greater than the C--S bond lengths of {\cmo}, but only slightly shorter than the C--Mo bond-length in $\gamma$-MoC.

Finally, we turn to {\cint}.
We find the energies of {\cintAboveS}, {\cintAboveMo} and {\cintBridge} (Fig.\,\ref{fig:geometry_structure_of_C_defects}(e-g)) to be 0.9, 0.4 and 0.5{\ev} respectively higher than {\cintMoC} (Fig.\,\ref{fig:geometry_structure_of_C_defects}(h)), and therefore conclude in agreement with previous modeling\,\cite{ataca-JPCC-115-13303} that {\cintMoC} is the equilibrium form of {\cint}.
We also note that geometry and energy depend somewhat upon the effective spin; the arrangements depicted in {\cintAboveS} and {\cintAboveMo} favor the spin-triplet configuration by 0.6 and 0.3{\ev} relative to spin-singlets, whereas {\cintBridge} and {\cintMoC} both favor the spin-singlet.
% removed sentence for repetition

However, we challenge the view that {\cint} is anything other than a metastable system.  
We find {\cint} displaces an S to form {\cs} bound to a sulfur interstitial, \cssi, as illustrated in Figs.\,\ref{fig:geometry_structure_of_C_defects}(i) and (j).
These two structures correspond to spin-triplet and spin-singlet electronic configurations, respectively, which we find to be indistinguishable in total energy. 
{\cssi} is around 1.7{\ev} lower in energy than the equilibrium form of {\cintMoC}.
%, and a preliminary estimate of the activation energy for the displacement of the S atom to form {\cssi} from {\cintMoC} is around 1.4{\ev}.
This stabilization of the kick-out of a sulfur atom 
%and the relatively low activation energy for the process to take place 
tends to mitigate against the proposal that {\cint} is responsible for the n-type behavior\,\cite{park-NPJ2DMA-7-60}.

\subsection{Energetics} 

In order to directly compare between C-containing centers of different compositions, one might consider formation energy, including underlying materials conditions (Mo- and S-rich conditions) as described in Sec.\,\ref{sec:method}. 

\begin{figure}[!htb]
 \begin{centering}
 \includegraphics[width=0.49\textwidth]{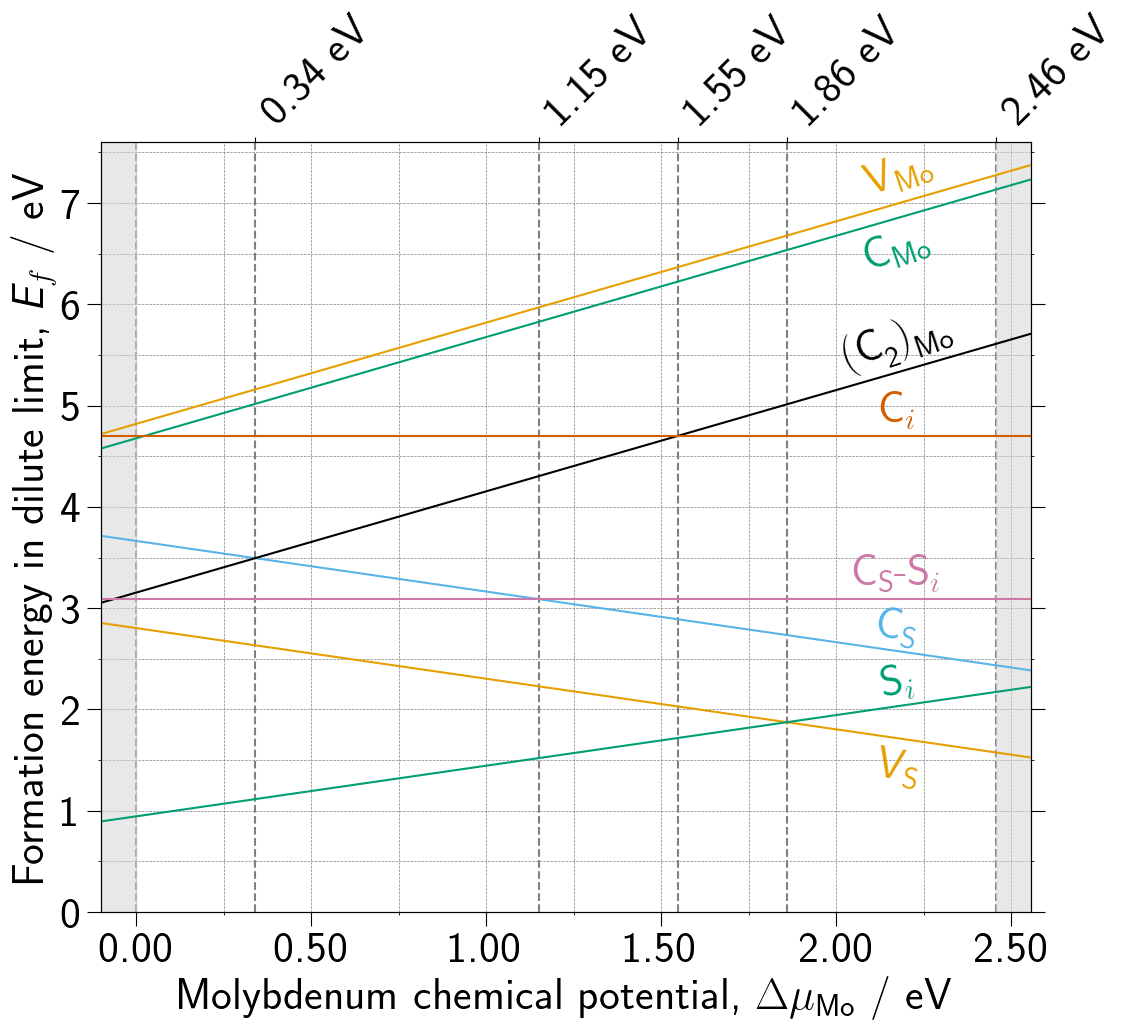}
 \caption{
 $E_f(\Delta\mu_{\rm Mo})$ for C-containing and native defects in monolayer {\mos}. 
 $\Delta\mu_{\rm Mo}$ is 0{\ev} and 2.46{\ev} for Mo-lean and Mo-rich limits. 
Dashed vertical lines highlight where $E_f$ for pairs of defects are equal. Energies based upon equilibrium forms in each case.}
 \label{Fig:MoChemPot_FormationEnergy_Graph}
 \end{centering}
\end{figure} 

\begin{table}[!htb]
\centering
\caption{$E_f\left(\Delta\mu_{\rm Mo}\right)$ and heats of reaction, $\Delta H$, for selected formation processes. Energies are quoted to the nearest 0.1{\ev}.}
\label{tab:FormationAndBindingEnergies}
\begin{tabular}{@{}ccr@{\,$\rightarrow$\,}l@{}}
\hline
Product & $E_f$ & \multicolumn{2}{c}{Reaction + $\Delta H$ } \\ \hline
{\cmo} & $4.7+\Delta\mu_{\rm Mo}$ & $\vmo+\cint$ & ${\cmo}+4.8$\\
{\cs} & $3.7-\Delta\mu_{\rm Mo}/2$ & $\vs+\cint$ & ${\cs}+3.8$ \\
{\ccmo} & $3.2+\Delta\mu_{\rm Mo}$ & $\cmo+\cint$ & ${\ccmo}+6.2$ \\
{\cssi} & $3.1$ & {\cintMoC} & ${\cssi}+1.7$\\ 
{\cssi} & $3.1$ & {\cs+\sint} & ${\cssi}+1.5$\\ 
\hline
{\cintMoC} & $4.8$ & \multicolumn{2}{c}{-} \\ 
{\vmo} & $4.8+\Delta\mu_{\rm Mo}$ & \multicolumn{2}{c}{-} \\
{\vs} & $2.8-\Delta\mu_{\rm Mo}/2$ & \multicolumn{2}{c}{-} \\ 
{\sint} & $0.9+\Delta\mu_{\rm Mo}/2$ & \multicolumn{2}{c}{-} \\ 
\hline
\end{tabular}
\end{table}

Fig.\,\ref{Fig:MoChemPot_FormationEnergy_Graph} shows a plot of $E_{f}\left(\Delta\mu_{\rm Mo}\right)$ and Table~\ref{tab:FormationAndBindingEnergies} lists the corresponding values. 
The data obtained are in line with the consensus in the literature (Sec.~\ref{sec:intro}) that {\vs} and {\sint} are relatively low energy point-defects in Mo-rich and S-rich limits, respectively.
The energies of C-containing centers are influenced by the choice of $\mu_{\rm C}$ (Sec.\,\ref{sec:method}), and these values should be compared with those of {\vs} and {\sint} with attention to this.

Of the carbon-containing centers, in the Mo-lean limit {\cssi}, {\ccmo} and {\cs} are most favorable, and therefore expected to be dominant under equilibrium conditions.
{\cmo} is comparatively high in formation energy, but remains significant as a potential precursor of {\ccmo}.
For Mo-rich conditions, {\cs} and {\cssi} are most energetically favorable. 
Our estimate of $E_f({\cs})$ is in close agreement with the literature\,\cite{park-NPJ2DMA-7-60}, providing additional support to our computational approach.

The relative energetic importance of different carbon centers may be crudely approximated using Boltzmann statistics, guided by typical growth temperatures\,\cite{Cho_Sim_Lee_Hoang_Seong_2023, Zhu_Shu_Jiang_Lv_Asokan_Omar_Yuan_Zhang_Jin_2017, Yue_Jian_Dong_Luo_Chang_2019}. 
Solubilities can be estimated using $\left[ X \right] \sim g \exp{\left( {E_f}/{k_{B}T}\right)}$, where $g$ is the site density, so that $[\vs]\sim[{\rm S}]\exp\left(-E_f(\vs)/k_BT\right)$ yields $[\vs]/[{\rm S}]\sim10^{-8}$ at {700\celcius} in S-lean conditions. 
Since {\vs} is more energetically favorable than other centers investigated, equilibrium concentrations of carbon defects would be expected to be lower.

\begin{table}[!htb]
\centering
\caption{
Adsorption energies, $E_{\rm ad}$, for various carbon interstitial sites calculated based upon different estimates of carbon atom energies, $E_{\rm C}$. 
For $E_{\rm ad}^{\rm coh}$, $E_{\rm C}$ is inferred from the cohesive energy of diamond (Sec.~\ref{sec:method}). $E_{\rm ad}^0$ and $E_{\rm ad}^1$ are obtained using $E_{\rm C}$ as the calculated energies of a free carbon atom in a spin singlet and spin triplet configuration, respectively, neglecting multiplet effects. Energies in eV.}
\label{tab:AdsorptionEnergies}
\newlength{\colspace}
\setlength{\colspace}{6pt}
\begin{tabular}{@{}l@{\hskip \colspace}|@{\hskip \colspace}c@{\hskip \colspace}c@{\hskip \colspace}c@{\hskip \colspace}|@{\hskip \colspace}c@{\hskip \colspace}c@{\hskip \colspace}c@{\hskip \colspace}c@{\hskip \colspace}c@{}}
% Old table config vv
%\begin{tabular}{@{}l|ccc|ccccc@{}}
\hline
&&&&\multicolumn{5}{c}{Literature values}\\
&$E_{\rm ad}^{\rm coh}$&$E_{\rm ad}^0$ & $E_{\rm ad}^1$ &\,\cite{liang-ACSAEM-2-1055} &\,\cite{li-AIPA-5} &\,\cite{ataca-JPCC-115-13303} &\,\cite{he-APL-96} &\,\cite{auni-ACSANM-28098}\\ \hline
{\cintMoC} & $2.7$ & $5.2$ & $3.6$ & & & $3.3$ & &\\
{\cintAboveMo} & $2.3$ & $4.8$ & $3.3$ & $4.7$ & $4.3$ & $2.7$ & $2.5$ & $3.4$ \\
{\cintAboveS} & $1.7$ & $4.2$ & $2.7$ & $3.8$ &  &  & & $2.8$ \\
{\cintAboveVoid} & $1.0$ & $3.4$ & $1.9$ & $3.3$ &  &  &  & $2.5$ \\
\hline
\end{tabular}
\end{table}

It is also insightful to evaluate the adhesion of adatoms to the surface. 
The adsorption energies, $E_{\rm ad}$, of {\cint} in various geometries are presented in Table\,\ref{tab:AdsorptionEnergies} alongside relevant literature data. 
As noted in Sec.~\ref{sec:method}, the reference energy of a free carbon atom is not trivial to obtain directly using DFT.
A na\"ive approach of obtaining the energy of an atom by simulating a single atom in a large periodic cell, populating the $2p$ orbitals equally with fractions of electrons will overestimate the atom energy, $E_{\rm C}$, and thus systematic overestimate $E_{\rm ad}$.
It seems likely that overestimates in the atom energy will be a significant contribution to the $>2.0${\ev} spread of $E_{\rm ad}$ indicated in Table~\ref{tab:AdsorptionEnergies}.

Using the empirically corrected atom energy, $E_{\rm ad}^{\rm coh}$, the adsorption of C onto {\mos} is exothermic, qualitatively consistent with the literature\,\cite{liang-ACSAEM-2-1055,li-AIPA-5,ataca-JPCC-115-13303,he-APL-96,auni-ACSANM-28098}, but quantitatively our predicted values are significantly lower. 
Subsequent desorption of C adatoms is therefore more likely than previously suggested, although it would remain less energetically favorable than chemically reacting with the {\mos} to produce {\cssi}.

$E_{\rm ad}$ has also been calculated for {\sint}.
Using the cohesive energy approach based upon cyclo-octosulfur\,\cite{brewer-BERK-1975} the adsorption of {\sint} onto {\mos} or {\cs} to form {\cssi} are estimated to be 1.9 and 3.4{\ev}, respectively.
The latter value suggests that adsorption of a C atom followed by displacement of sulfur is unlikely to result in subsequent desorption of the sulfur intersticy.

Finally, we consider the reaction energetics as a guide to thermal stability, utilizing calculated heats of reaction and energy barriers.
Five reactions key to the incorporation of carbon in {\mos} are listed in Table~\ref{tab:FormationAndBindingEnergies}, all of which are found to be exothermic.

Several of these reactions involve the diffusion of {\cint}. Its minimum energy diffusion barrier is via the {\cintAboveMo} site and is estimated at 1.1{\ev}, which is competitive in comparison with {\sint} diffusion calculated at 1.8{\ev} (in agreement with \,\cite{komsa-PRB-91-125304,gusakov-PSSB-259-2100479}).
For processes driven by the addition of interstitial species, one can imagine two chains emerging, incrementally increasing thermodynamic stability: $\vmo\xrightarrow{{\rm C}_i}\cmo\xrightarrow{{\rm C}_i}\ccmo$ and $\vs\xrightarrow{{\rm C}_i}\cs\xrightarrow{{\rm S}_i}\cssi$.
Formation of {\ccmo}, {\cs} and {\cssi} during growth under Mo-lean conditions may be possible by this incremental capture of interstitials, as each step is exothermic.
Since the reaction energies are several eV, reverse reactions are energetically unfavorable. 
In particular, the kick-out process ${\cint} \rightarrow {\cssi}$ has a forward barrier of 1.4{\ev} and 3.1{\ev} in reverse, while the barrier to the binding of ${\cs}+{\sint} \rightarrow {\cssi}$ is calculated at 1.8{\ev}, and 3.4{\ev} in reverse.
Unless the complexes and substitutional centers are able to diffuse as units, these are the point defects that are amongst the candidates most likely to be present in sufficient concentrations for observation by spectroscopic means.
Since {\ccmo} is four times more strongly bound than {\cssi}, we suggest a focus should therefore be upon the carbon pair.

One may consider {\cs} formation from {\vs} via the interim formation of {\cssi}: {\vs}\,+\,{\cint}\,\,$\rightarrow$\,{\vs}\,+\,{\cssi}\,$\rightarrow$\,{\vs}\,+\,{\cs}\,+\,{\sint}\,$\rightarrow$\,{\cs}. This route involves {\sint} diffusion as opposed to {\cint}, however, since the barrier to dissociation of {\sint} from {\cssi} is 3.4{\ev} it is unlikely to occur.
%Although the diffusion of {\cint} is preferred to {\sint} indicated by its lower barrier, {\cint} diffusion will cease once  {\cint} has converted to {\cssi} by the kick-out process.

\subsection{Electronic structures\label{sec:bandstructure}}

\begin{figure*}[t!]
\begin{minipage}{0.129\textwidth}\includegraphics[width=\textwidth]{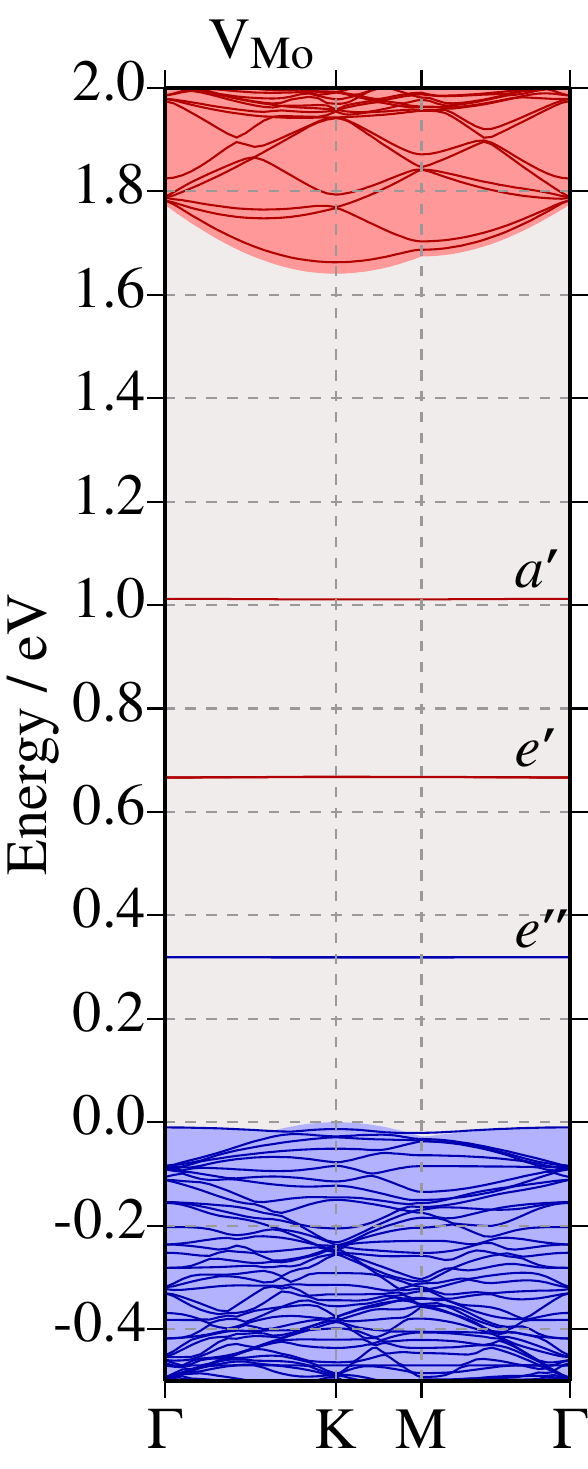}\end{minipage}%
\begin{minipage}{0.0967\textwidth}\includegraphics[width=\textwidth]{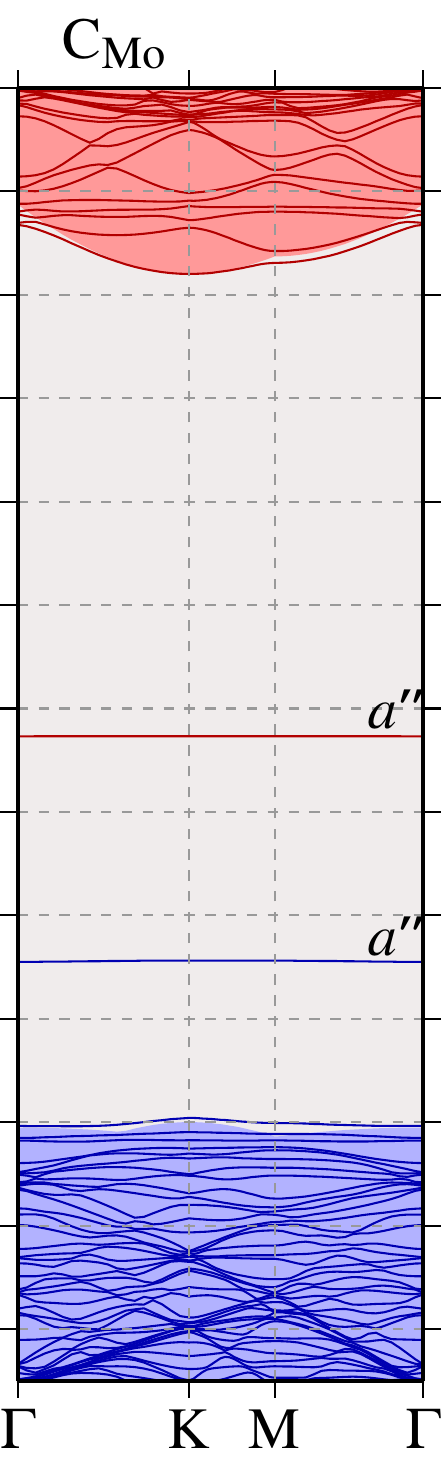}\end{minipage}%
\begin{minipage}{0.0967\textwidth}\includegraphics[width=\textwidth]{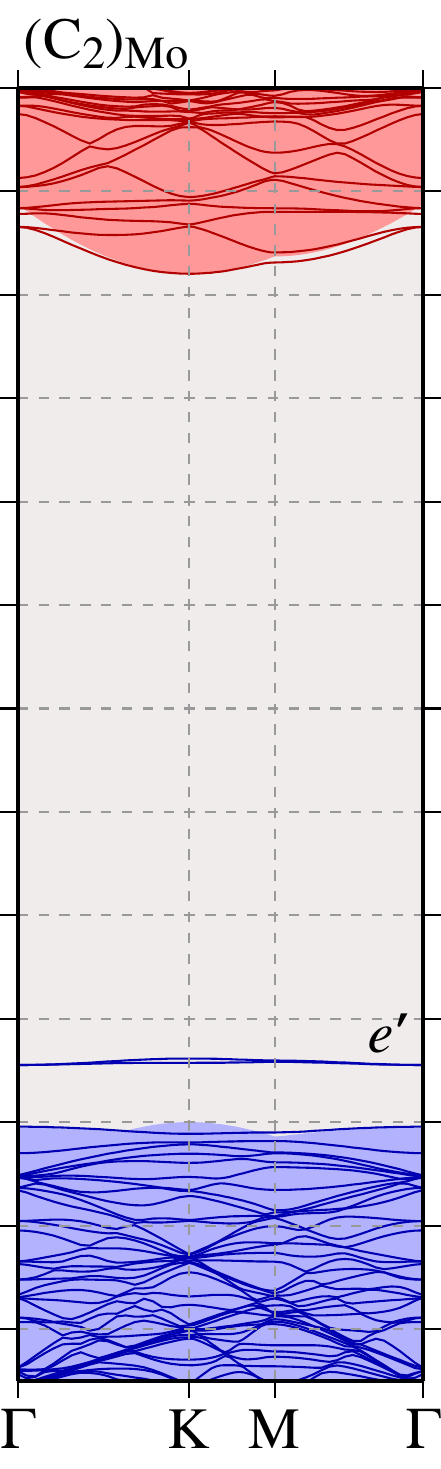}\end{minipage}%
\begin{minipage}{0.0967\textwidth}\includegraphics[width=\textwidth]{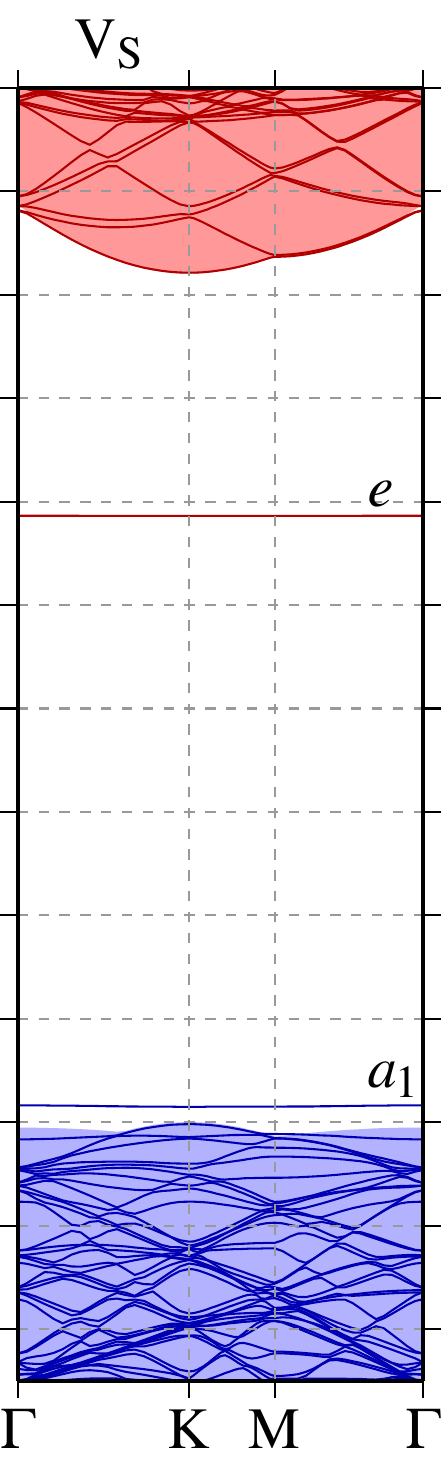}\end{minipage}%
\begin{minipage}{0.0967\textwidth}\includegraphics[width=\textwidth]{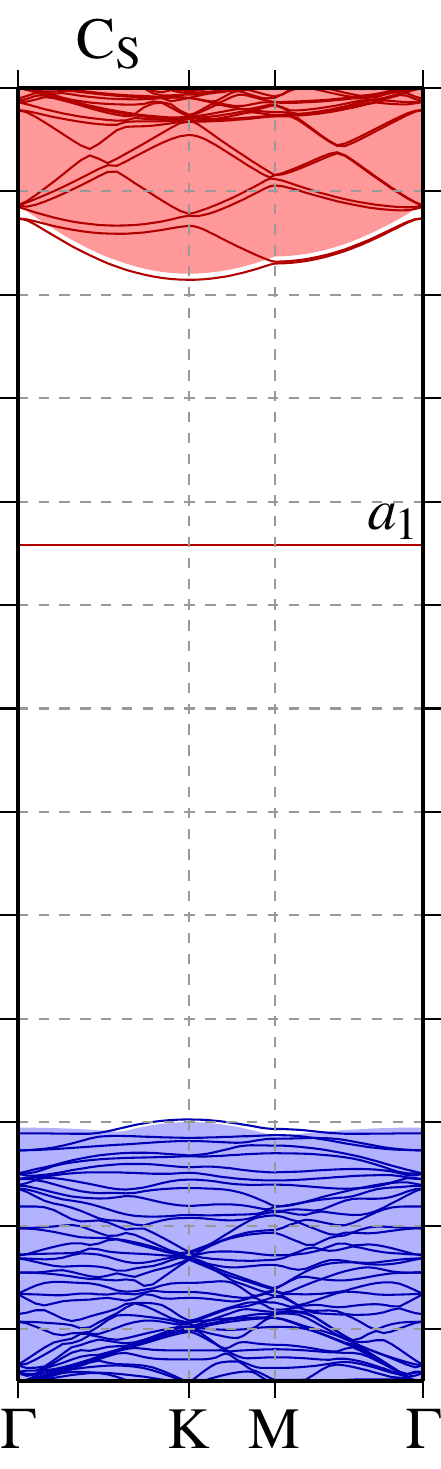}\end{minipage}%
\begin{minipage}{0.0967\textwidth}\includegraphics[width=\textwidth]{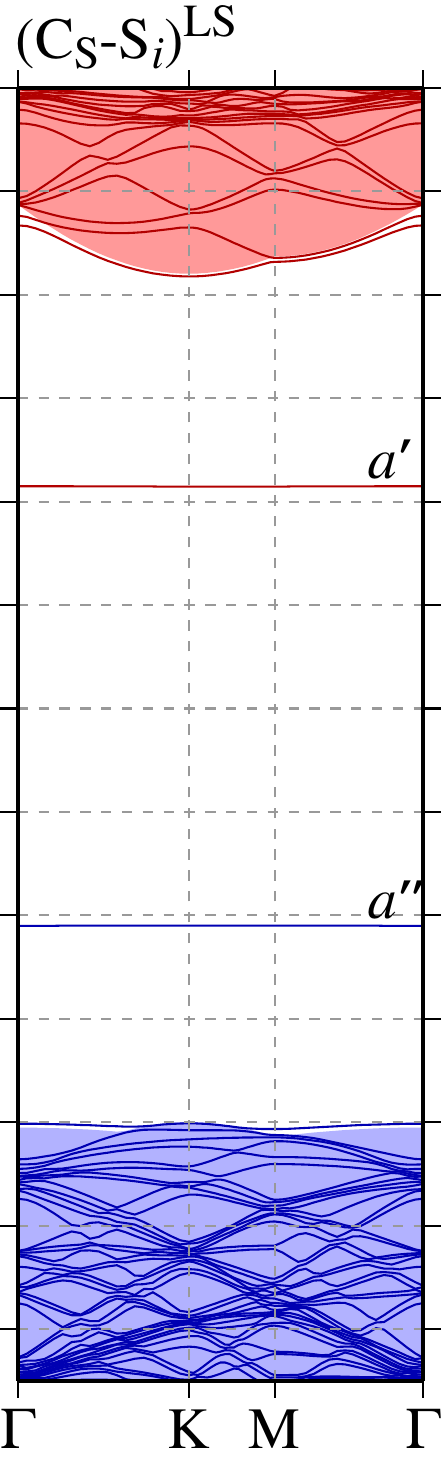}\end{minipage}%
\begin{minipage}{0.0967\textwidth}\includegraphics[width=\textwidth]{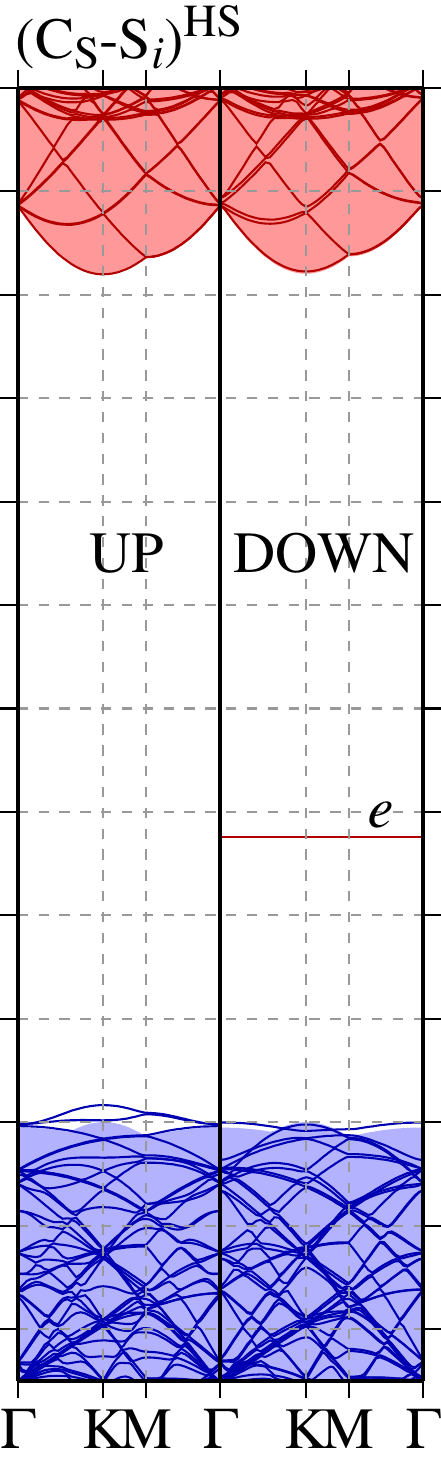}\end{minipage}%
\begin{minipage}{0.0967\textwidth}\includegraphics[width=\textwidth]{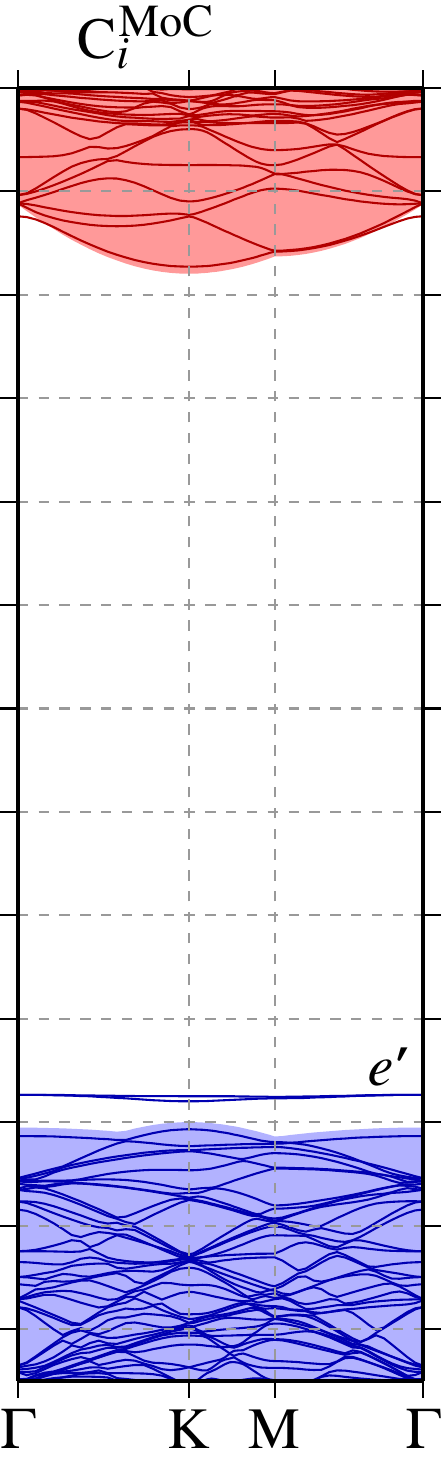}\end{minipage}%
\begin{minipage}{0.0967\textwidth}\includegraphics[width=\textwidth]{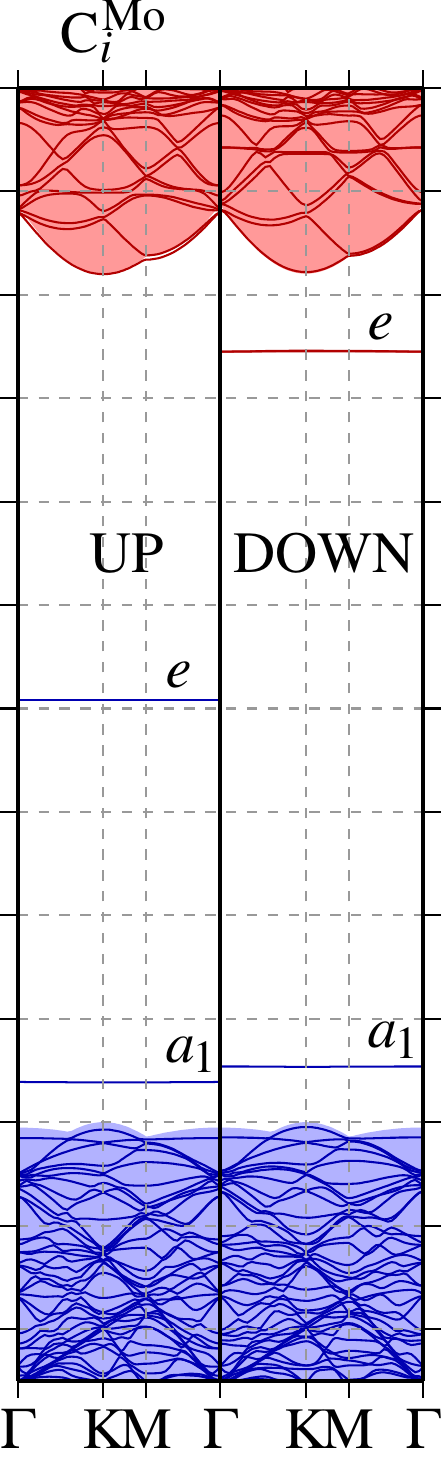}\end{minipage}%
\begin{minipage}{0.0967\textwidth}\includegraphics[width=\textwidth]{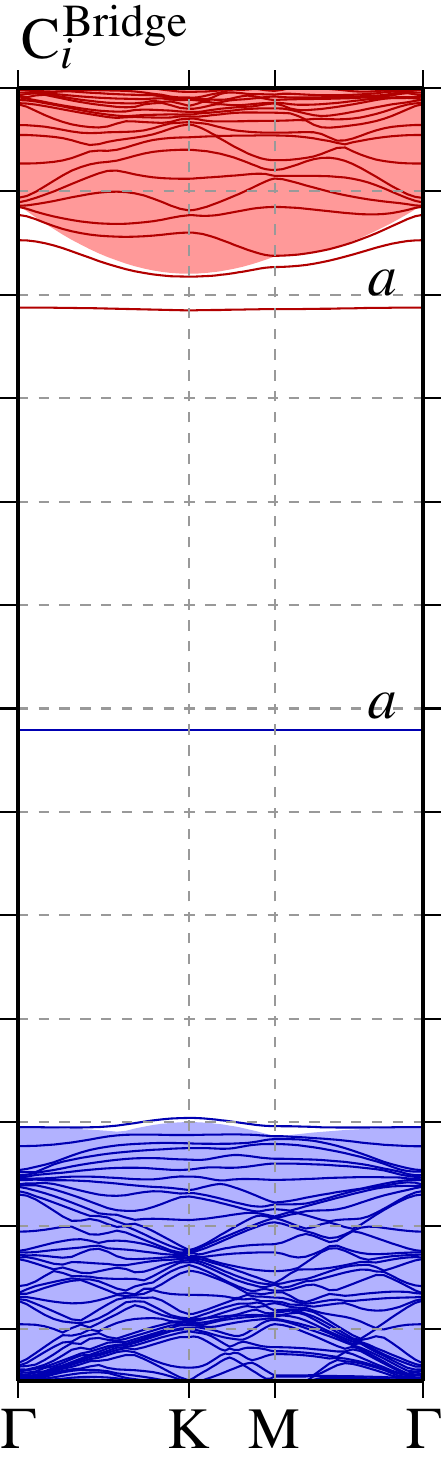}\end{minipage}%

%\begin{minipage}{0.13\textwidth}\includegraphics[width=\textwidth]{Bandst/CSandSiLS.eps}\end{minipage}%
%\begin{minipage}{0.13\textwidth}\includegraphics[width=\textwidth]{Bandst/CSandSiHS.eps}\end{minipage}%
 \caption{Electronic band structure in the vicinity of the band-gap for selected defects in monolayer {\mos}. 
 Occupied(empty) bands are shown in red(blue), with shading showing envelopes of the defect-free valence and conduction bands.
 Zero on the energy scale is the valence band maximum.
 Band structures are spin-averaged except for ({\cssi})$^{\rm HS}$ which is the spin-triplet, with left and right panels showing up and down spin channels, respectively.
 Gap-state labels show irreducible representations at $\Gamma$, and  {\vmo} and {\vs} have been included for comparison. 
 }
\label{fig:Bandstructures}
\end{figure*}

% Native defect comparison with literature vv
% Vs xu-MTC-22-100772 zhou-NL-13-2615 feng-JAC-122 feng-MCP-146
% VMo feng-JAC-122 feng-MCP-146
% Carbon impurity comparison with literature vv
% Cs park-NPJ2DMA-7-60 auni-ACSANM-28098
% Ci ataca-JPCC-115-13303
% CiAboveMo ataca-JPCC-115-13303 park-NPJ2DMA-7-60 auni-ACSANM-28098
%The systems presented in this comparative literature are smaller, around the $3\times3$ to $4\times4$ supercell range, leading to a greater dispersion of the localized states. Furthermore, comparison of the defect gap states in monolayer {\mos} to the analogous C defects in the bulk material presented by Ref.\,\cite{park-NPJ2DMA-7-60} is possible by accounting for the bulk material's elevated energy of the VBM at $\Gamma$ relative to monolayer. 

Next we turn to electronic structure, depicted in Fig.\,\ref{fig:Bandstructures}, and initially focus upon substitutional centers.
Defect-induced band-gap states may act as carrier sources and/or traps, optical transition mechanisms and carrier scattering centers in devices and hence impact optoelectronic performance.
Vacancy band structures shown in Fig.\,\ref{fig:Bandstructures} provide benchmark data, and are in good agreement with the literature\,\cite{xu-MTC-22-100772,zhou-NL-13-2615,feng-JAC-122,feng-MCP-146}, giving confidence to our calculated electronic structures.

{\cmo} and {\ccmo} both give rise to deep gap-states. 
The unsaturated S-sites in {\cmo}, just as with {\vmo}, result in both occupied and empty states, whereas the fully bonded {\ccmo} only results in occupied states in the vicinity of the valence band maximum.
The relatively low symmetry of {\cmo} and fewer unsaturated sites means that there are fewer gap states than {\vmo}.
However, the depth of the levels suggests that {\cmo} would be able to trap electrons and possibly holes, which is understood to be the case for {\vmo}\,\cite{feng-JAC-122}.
The presence of both occupied and empty states in the gap for {\cmo} also gives rise to the potential of an optical transition. 
Based upon the band structure, the zero phonon optical absorption energies would be around 0.5{\ev}, rendering it in the near infra-red.
In contrast, the band structure of {\ccmo} suggests it to be both electrically inactive, and although transitions involving localized states resonant with the host bands cannot be ruled out, it seems unlikely {\ccmo} would give rise to sharp optical transitions, there being only occupied levels very close to the valence band maximum.

{\cs}, amongst other carbon-containing centers, has been been suggested to lead to p-type doping\,\cite{park-NPJ2DMA-7-60}.
Fig.\,\ref{fig:Bandstructures} shows a single localized, unoccupied state lying above mid-gap.
This is also what is seen in the band diagrams in Refs.\,\cite{park-NPJ2DMA-7-60} and\,\cite{auni-ACSANM-28098}, however, the gap-state is self-evidently a deep acceptor level, inconsistent with the assertion that {\cs} could be responsible for p-type conduction.
A crude estimate of the acceptor level would place it at more than 1{\ev} above the valence band, so that thermal ionization would be energetically unfavorable at room temperature.

%\begin{figure}[t!]
%\begin{minipage}{0.116\textwidth}\includegraphics[width=\textwidth]{Bandst/CiMoC.eps}\end{minipage}%
%\begin{minipage}{0.087\textwidth}\includegraphics[width=\textwidth]{Bandst/CiAboveMo.eps}\end{minipage}%
%\begin{minipage}{0.087\textwidth}\includegraphics[width=\textwidth]{Bandst/Cibridge.eps}\end{minipage}%
%\begin{minipage}{0.087\textwidth}\includegraphics[width=\textwidth]{Bandst/CSandSiLS.eps}\end{minipage}%
%\begin{minipage}{0.087\textwidth}\includegraphics[width=\textwidth]{Bandst/CSandSiHS.eps}\end{minipage}%
% \caption{Electronic band structure in the vicinity of the band-gap for selected interstitial centers of monolayer {\mos}.
% Coloring and labels are as in Fig.~\ref{fig:Bandstructures}.}
%\label{fig:Bandstructures-Interstices}
%\end{figure}

Next, we review the electronic structure of the interstitial centers, which includes non-equilibrium arrangements for comparison with the literature.
All structures examined give rise to gap levels.
The pair of filled states close to the valence band edge for {\cintMoC} are also reported in Ref.\,\cite{ataca-JPCC-115-13303}. 
The gap states of {\cintAboveMo} and {\cintBridge} are also in agreement with the literature\,\cite{auni-ACSANM-28098,park-NPJ2DMA-7-60,ataca-JPCC-115-13303}.

The fact that there is only a single occupied, degenerate band {\cintMoC} close to the valence band maximum is consistent with this being a low energy structure, with other, higher energy forms of {\cint}, including {\cintAboveVoid} (carbon in the hexagonal void, in-plane with sulfurs) and {\cintAboveS} (neither are shown) possessing bands deep in the band-gap.
The equilibrium interstitial structure, in contrast to the other arrangements, does not present obvious optical transition mechanisms, other than those involving the underlying {\mos} bands, such as photo-ionization.

Crucially, neither the interstitial structures present in Ref.\,\cite{park-NPJ2DMA-7-60}, nor {\cintMoC} in its most stable configuration present shallow donor levels to give rise to n-type behavior at room temperature, as they report.
We conclude that C interstitial centers are not the cause of n-type conductivity.

{\cssi} is compositionally identical to {\cint}, but significantly lower in energy.
Spin singlet and triplet states have different structures, and correspondingly different band structures.
These centers are more favorable than {\cint} and therefore more likely to be observed in experiment.
Based upon occupancy, electron-spin conservation and dipole selection rules, there are transition mechanisms for the spin-singlet form, ({\cssi})$^{\rm LS}$.
Based purely upon the band structure, the zero phonon optical absorption energies for these systems would be around 0.9{\ev}, being in the near infrared range. 
The spin-triplet band structure does not indicate an inter-level transition, and neither form is an obvious candidate for releasing either electrons or holes to generate electrical conduction.

Indeed, we conclude that the electronic structures for all carbon-containing centers examined indicate that it will only act as a deep carrier trap, suggesting it would impede electrical conduction in the absence of other components (such as hydrogen~\cite{singh-PRB-121201}).

\subsection{Vibrational Modes \label{Section:Results-VM}}

\begin{table*}[!htb]
\centering
\caption{Local vibrational modes of carbon defects in monolayer {\mos}, their point-group symmetry and irreducible representations. Local modes are identified as lying above 456{\cm}, the calculated one-phonon maximum. All modes listed are infrared and Raman active except $\omega_6$ which is Raman-inactive and $\omega_7$ and $\omega_8$ which are IR-inactive. Wave numbers are provided for $^{12}$C, with the shift to lower wave number for $^{13}$C shown in parentheses. For {\ccmo}, the two shifts correspond to first the mixed-isotope case and then $^{13}$C, with the mixed-isotope case possessing $C_{3v}$ symmetry.}
\label{tab:VMs}
\begin{tabular}{ccccl}
\hline
Defect & Label & {Sym.} & Wave number & {Description of predominant mode contributions}\\ \hline
{\cs} & {$\omega_{0}$} & $C_{3v}$, $A_1$ & {578(19)} & {C--Mo--S bend/scissor.} \\
& {$\omega_{1}$} & $C_{3v}$, $E$ & {638(23)} & {C--Mo stretch.} \\ \hline
{\cmo}& {$\omega_{2}$} & $C_{1h}$, $A^{\prime\prime}$ & 641(32) & {C--S stretch in {\mos} plane.} \\
& {$\omega_{3}$} & $C_{1h}$, {${A}^{\prime}$} & 680(22)& {C--S stretch in {\mos} plane.} \\
& {$\omega_{4}$} & $C_{1h}$, {${A}^{\prime}$} & 722(24) & {C--S stretch normal to {\mos} plane.} \\ \hline
{\ccmo}
& {$\omega_{5}$} & $D_{3h}$, {${E}^{\prime}$} & 675(13, 22)&{In-phase stretch of two C--S.} \\% & {Y} & {Y} & {675} & {662} & {653} \\
& {$\omega_{6}$} & $D_{3h}$, {${A}^{\prime}_2$} & 693(11, 20)&{In-phase bending of all C--S bonds.} \\%& {Y} & {N} & {693} & {682} & {673} \\
 & {$\omega_{7}$} & $D_{3h}$, {${E}^{\prime \prime}$} & 782(13, 29) &{The corresponding anti-phase C--S stretch to $\omega_5.$} \\%& {N} & {Y} & {782} & {769} & {753} \\
 & {$\omega_{8}$} & $D_{3h}$, {${A}^{\prime}_1$} & 935(17, 36) &{C--C stretch.}\\% & {N} & {Y} & {935} & {918} & {899} \\ 
%\hline {\cssi} ($S=1$) & {$\omega_{9}$} & $C_{3v}$, {$E$} & ?? & {In-plane oscillation of the carbon.}\\% & {Y} & {Y} & \multicolumn{3}{c}{{Mixes with phonon modes, discussed in text.}} \\
\hline ({\cssi})$^{\rm HS}$ & {$\omega_{10}$} & $C_{3v}$, {${A}_1$} & 971(29)& {C--{\sint} stretch.}\\% & {Y} & {Y} & {971} & {942} & {-} \\ 
\hline ({\cssi})$^{\rm LS}$ & {$\omega_{11}$} & $C_{1h}$, {${A}^{\prime}$} & 499(18)& {C--Mo stretch.}\\% & {Y} & {Y} & {499} & {481} & {-} \\
& {$\omega_{12}$} & $C_{1h}$, {${A}^{\prime \prime}$} & 540(19)& {Anti-phase stretch of two C--Mo.}\\% & {Y} & {Y} & {540} & {521} & {-} \\
& {$\omega_{13}$} & $C_{1h}$, {${A}^{\prime}$} & 971(31)& {C--{\sint} stretch.}\\% & {Y} & {Y} & {971} & {940} & {-} \\
\hline 
\end{tabular}
\end{table*}

The final defect properties we have evaluated are the local vibrational modes (LVMs), which are listed in Table\,\ref{tab:VMs}. 
In the absence of comparative vibrational data for C impurities, we have calculated the vibrational modes of ethanethiol, {CH}$_3${CH}$_2${SH} to provide a benchmark.
C--S and C--C stretch modes are calculated at 647 and 957{\cm}, respectively, compared to 663 and 969{\cm} from experiment\cite{wolff-CJC-63-1708}.
We take this approx{.} {3\%} underestimate to be indicative of the level of accuracy of the predicted LVMs listed in Table\,\ref{tab:VMs}.

We note the C--S and C--C stretch modes ($\omega_5$--$\omega_8$) are in the same spectral range as ethanethiol stretch modes. 
Crucially, all defects investigated possesses at least one LVM that would be observable in IR and/or Raman spectroscopy, and when combined with isotopic shifts and polarization, these modes would be expected to provide a high degree of specificity for determining the microstructure.
This may be particularly important for {\cs},{\cssi}$^{\rm HS}$ and {\ccmo}, as there is little prospect of detecting these centers via electronic transitions due to the absence of optical transition mechanisms in these cases (see Sec.\,\ref{sec:bandstructure}).

\section{Summary}

An exploration of carbon point defects in monolayer {\mos} from first principles has been conducted. 
Impurities at the sulfur and molybdenum sites have been investigated, in addition to the carbon interstitial guided by the proposal that {\cint} might lead to electrical doping and associated conductivity\,\cite{park-NPJ2DMA-7-60}. 

The first conclusion we draw is that the literature proposals\,\cite{liang-ACSAEM-2-1055,park-NPJ2DMA-7-60,auni-ACSANM-28098,he-APL-96,li-AIPA-5} for the geometry of both {\cmo} and {\cint} should be reconsidered.
In the case of {\cmo}, chemical reconstruction to form a four-fold covalently bonding carbon species is 0.6{\ev} lower in energy than the three-fold co-ordinated structure resulting from an off-site displacement of carbon along [0001].
Numerous geometries have been proposed in the literature for {\cint}, predominantly a structure featuring carbon co-ordinated with sulfur.
However, we confirm that the the equilibrium geometry identified in an earlier publication\,\cite{ataca-JPCC-115-13303} is $\sim$0.5{\ev} lower in energy than the commonly reported geometry of {\cintAboveMo}. 
The favorability of {\cintMoC} suggests both that carbon would be able to move through {\mos} layers due to its intermediate vertical position in the {\mos} layer, and that the molybdenum carbide bond formation is energetically favorable despite the apparent limited space to accommodate the carbon interstitial.
Critically, none of the the {\cint} centers modeled are realistic candidates for electrical of doping of {\mos}, as they give rise to deep band-gap states. 
This strongly counters the suggestion that {\cint} is responsible for observed electrical conductivity~\cite{park-NPJ2DMA-7-60}.

Indeed, it is not plausible from our data to conceive that any of the carbon or native defects modeled here are likely to be a source of free carriers, and therefore they are unable to account for the evolution from p-type to n-type by the addition of carbon as suggested previously\,\cite{park-NPJ2DMA-7-60}.
In particular, sulfur vacancies are deep acceptors, and unable to introduce holes to the valence band under ambient conditions.
Interstitial carbon in any of the many forms reviewed, including the compositionally identical {\cssi}, possess occupied defect bands only in the lower part of the band gap, and are therefore unable to introduce electrons to the conduction band under ambient conditions.
We conclude that the experimental observations of n-type conduction\,\cite{park-NPJ2DMA-7-60} remain entirely unexplained by carbon or monovacancy defects.

The second area of focus is the identification of two novel stable structures, {\ccmo} and {\cssi}, which to our knowledge have not previously been considered. 
These centers represent the energetically favorable carbon-containing point defects of carbon at the molybdenum and sulfur sites, respectively.

Thirdly, when considering the centers likely to be present in {\mos} when contaminated by carbon, we focus upon the formation energies in the Mo- and S-rich limits.
{\cssi} and {\ccmo} are energetically most favorable under S-rich growth conditions, whereas under Mo-rich conditions, the most favorable carbon point defect is {\cs}.

Finally, we have provided vibrational wavenumbers for local modes that provide a potential route to unambiguous identification of the microstructures of carbon-containing defects in {\mos}.
% Where carbon is present in the material, it is of great value to have specific observable properties that uniquely identify the microstructure.
% Band structure and vibrational data provide insight into their potential for optical detection (including electronic transitions visible in luminescence and vibrational signatures visible in infrared absorption and/or Raman scattering), and to assess their potential impact on charge carriers due to trapping or emission.
% All carbon defects investigated defects introduce localized mid-gap electronic states and are thus potential carrier traps and scattering centers, but only {\cssi}, {\cmo} and higher energy structures of {\cint} introduce both occupied and empty states with dipole-allowed electronic transitions. 
% Of course, optical and electrical transitions may be either beneficial or detrimental, depending upon the device application. 
% For example, carbon contamination is likely to be undesirable for photovoltaic applications due to non-radiative recombination and carrier mobility reduction, but optical transition may have utility in the formation of single-photon emitters, valued for quantum technology applications. 
% All investigated defects introduce at least one IR/Raman active LVM above the one phonon maximum, but with the most favorable formation energies lying in the 2--4{\ev} range, it is unclear whether the concentrations in which carbon might be taken up would be sufficient for direct detection of these LVMs.
The direct, spectroscopic identification of carbon microstructure is critical to resolving what, if any, role carbon point-defects have on the production of n-type conduction in {\mos}.

% Overall, it is suspected that carbon impurities on the sulfur site will detriment opto-electronic performance to a greater degree than those at the molybdenum site on account of their prevalence: the formation energy of {\vs} (from which they can be derived) is the lowest of the native defects, the formation energy of {\cs} is the lowest of the carbon impurities in the Mo-rich limit, whereas {\cssi} is the lowest in the Mo-lean limit. Taking these factors in conjunction with a comparable effect upon opto-electronic performance between carbon impurities at the Mo-site and S-site due to the introduction of a comparable number of mid-gap electronic states, it is clear that the prevalence of S-site defects will cause more opto-electronic adverse impact than those at the Mo-site.

In summary, we present an expanded picture of carbon impurities in monolayer {\mos}, increasing not only in breadth of defects considered, but also in details of derived quantities and observables. 
These data are intended to facilitate the experimental exploration of carbon impurities.

\section{Acknowledgments}

%Authors gratefully acknowledge Prof. Patrick Briddon and Dr. Mark Rayson for use of AIMPRO. 
The authors acknowledge the Engineering and Physical Sciences Research Council (EPSRC) Centre for Doctoral Training in Renewable Energy Northeast Universities (ReNU) for funding through grant EP/S023836/1.
FA and JG also gratefully acknowledge the support provided by the Najran University and the Saudi Arabian Ministry of Education in facilitating this research endeavor.

\bibliographystyle{apsrev4-2}
\bibliography{submissionDraft}

\begin{thebibliography}{74}%
\makeatletter
\providecommand \@ifxundefined [1]{%
 \@ifx{#1\undefined}
}%
\providecommand \@ifnum [1]{%
 \ifnum #1\expandafter \@firstoftwo
 \else \expandafter \@secondoftwo
 \fi
}%
\providecommand \@ifx [1]{%
 \ifx #1\expandafter \@firstoftwo
 \else \expandafter \@secondoftwo
 \fi
}%
\providecommand \natexlab [1]{#1}%
\providecommand \enquote  [1]{``#1''}%
\providecommand \bibnamefont  [1]{#1}%
\providecommand \bibfnamefont [1]{#1}%
\providecommand \citenamefont [1]{#1}%
\providecommand \href@noop [0]{\@secondoftwo}%
\providecommand \href [0]{\begingroup \@sanitize@url \@href}%
\providecommand \@href[1]{\@@startlink{#1}\@@href}%
\providecommand \@@href[1]{\endgroup#1\@@endlink}%
\providecommand \@sanitize@url [0]{\catcode `\\12\catcode `\$12\catcode `\&12\catcode `\#12\catcode `\^12\catcode `\_12\catcode `\%12\relax}%
\providecommand \@@startlink[1]{}%
\providecommand \@@endlink[0]{}%
\providecommand \url  [0]{\begingroup\@sanitize@url \@url }%
\providecommand \@url [1]{\endgroup\@href {#1}{\urlprefix }}%
\providecommand \urlprefix  [0]{URL }%
\providecommand \Eprint [0]{\href }%
\providecommand \doibase [0]{https://doi.org/}%
\providecommand \selectlanguage [0]{\@gobble}%
\providecommand \bibinfo  [0]{\@secondoftwo}%
\providecommand \bibfield  [0]{\@secondoftwo}%
\providecommand \translation [1]{[#1]}%
\providecommand \BibitemOpen [0]{}%
\providecommand \bibitemStop [0]{}%
\providecommand \bibitemNoStop [0]{.\EOS\space}%
\providecommand \EOS [0]{\spacefactor3000\relax}%
\providecommand \BibitemShut  [1]{\csname bibitem#1\endcsname}%
\let\auto@bib@innerbib\@empty
%</preamble>
\bibitem [{\citenamefont {{Novoselov}}\ \emph {et~al.}(2004)\citenamefont {{Novoselov}}, \citenamefont {{Geim}}, \citenamefont {{Morozov}}, \citenamefont {{Jiang}}, \citenamefont {{Zhang}}, \citenamefont {{Dubonos}}, \citenamefont {{Grigorieva}},\ and\ \citenamefont {{Firsov}}}]{novoselov-S-306-666}%
  \BibitemOpen
  \bibfield  {author} {\bibinfo {author} {\bibfnamefont {K.~S.}\ \bibnamefont {{Novoselov}}}, \bibinfo {author} {\bibfnamefont {A.~K.}\ \bibnamefont {{Geim}}}, \bibinfo {author} {\bibfnamefont {S.~V.}\ \bibnamefont {{Morozov}}}, \bibinfo {author} {\bibfnamefont {D.}~\bibnamefont {{Jiang}}}, \bibinfo {author} {\bibfnamefont {Y.}~\bibnamefont {{Zhang}}}, \bibinfo {author} {\bibfnamefont {S.~V.}\ \bibnamefont {{Dubonos}}}, \bibinfo {author} {\bibfnamefont {I.~V.}\ \bibnamefont {{Grigorieva}}},\ and\ \bibinfo {author} {\bibfnamefont {A.~A.}\ \bibnamefont {{Firsov}}},\ }\href {https://doi.org/10.1126/science.1102896} {\bibfield  {journal} {\bibinfo  {journal} {Science}\ }\textbf {\bibinfo {volume} {306}},\ \bibinfo {pages} {666–669} (\bibinfo {year} {2004})}\BibitemShut {NoStop}%
\bibitem [{\citenamefont {Frank}\ \emph {et~al.}(2001)\citenamefont {Frank}, \citenamefont {Dennard}, \citenamefont {Nowak}, \citenamefont {Solomon}, \citenamefont {Taur},\ and\ \citenamefont {Wong}}]{frank-PIEEE-259}%
  \BibitemOpen
  \bibfield  {author} {\bibinfo {author} {\bibfnamefont {D.}~\bibnamefont {Frank}}, \bibinfo {author} {\bibfnamefont {R.}~\bibnamefont {Dennard}}, \bibinfo {author} {\bibfnamefont {E.}~\bibnamefont {Nowak}}, \bibinfo {author} {\bibfnamefont {P.}~\bibnamefont {Solomon}}, \bibinfo {author} {\bibfnamefont {Y.}~\bibnamefont {Taur}},\ and\ \bibinfo {author} {\bibfnamefont {H.-S.~P.}\ \bibnamefont {Wong}},\ }\href {https://doi.org/10.1109/5.915374} {\bibfield  {journal} {\bibinfo  {journal} {Proc{.} IEEE}\ }\textbf {\bibinfo {volume} {89}},\ \bibinfo {pages} {259} (\bibinfo {year} {2001})}\BibitemShut {NoStop}%
\bibitem [{\citenamefont {Cong}\ \emph {et~al.}(2015)\citenamefont {Cong}, \citenamefont {Tang}, \citenamefont {Zhao},\ and\ \citenamefont {Chu}}]{cong-SR-5-9361}%
  \BibitemOpen
  \bibfield  {author} {\bibinfo {author} {\bibfnamefont {W.~T.}\ \bibnamefont {Cong}}, \bibinfo {author} {\bibfnamefont {Z.}~\bibnamefont {Tang}}, \bibinfo {author} {\bibfnamefont {X.~G.}\ \bibnamefont {Zhao}},\ and\ \bibinfo {author} {\bibfnamefont {J.~H.}\ \bibnamefont {Chu}},\ }\href {https://doi.org/10.1038/srep09361} {\bibfield  {journal} {\bibinfo  {journal} {Scientific Reports}\ }\textbf {\bibinfo {volume} {5}},\ \bibinfo {pages} {9361} (\bibinfo {year} {2015})}\BibitemShut {NoStop}%
\bibitem [{\citenamefont {Li}\ \emph {et~al.}(2018)\citenamefont {Li}, \citenamefont {Cao}, \citenamefont {Wang}, \citenamefont {Xiao}, \citenamefont {Li}, \citenamefont {Delaunay},\ and\ \citenamefont {Zhu}}]{li-CSR-47-4981}%
  \BibitemOpen
  \bibfield  {author} {\bibinfo {author} {\bibfnamefont {C.}~\bibnamefont {Li}}, \bibinfo {author} {\bibfnamefont {Q.}~\bibnamefont {Cao}}, \bibinfo {author} {\bibfnamefont {F.}~\bibnamefont {Wang}}, \bibinfo {author} {\bibfnamefont {Y.}~\bibnamefont {Xiao}}, \bibinfo {author} {\bibfnamefont {Y.}~\bibnamefont {Li}}, \bibinfo {author} {\bibfnamefont {J.-J.}\ \bibnamefont {Delaunay}},\ and\ \bibinfo {author} {\bibfnamefont {H.}~\bibnamefont {Zhu}},\ }\href {https://doi.org/10.1039/c8cs00067k} {\bibfield  {journal} {\bibinfo  {journal} {Chem{.} Soc{.} Rev{.}}\ }\textbf {\bibinfo {volume} {47}},\ \bibinfo {pages} {4981–5037} (\bibinfo {year} {2018})}\BibitemShut {NoStop}%
\bibitem [{\citenamefont {Yin}\ \emph {et~al.}(2022)\citenamefont {Yin}, \citenamefont {Zhang}, \citenamefont {Shao}, \citenamefont {Robertson},\ and\ \citenamefont {Guo}}]{yin-NPJ2DMA-2397}%
  \BibitemOpen
  \bibfield  {author} {\bibinfo {author} {\bibfnamefont {Y.}~\bibnamefont {Yin}}, \bibinfo {author} {\bibfnamefont {Z.}~\bibnamefont {Zhang}}, \bibinfo {author} {\bibfnamefont {C.}~\bibnamefont {Shao}}, \bibinfo {author} {\bibfnamefont {J.}~\bibnamefont {Robertson}},\ and\ \bibinfo {author} {\bibfnamefont {Y.}~\bibnamefont {Guo}},\ }\href {https://doi.org/10.1038/s41699-022-00332-6} {\bibfield  {journal} {\bibinfo  {journal} {npj 2D Mat{.} and Appl{.}}\ }\textbf {\bibinfo {volume} {6}},\ \bibinfo {pages} {1} (\bibinfo {year} {2022})}\BibitemShut {NoStop}%
\bibitem [{\citenamefont {Radisavljevic}\ \emph {et~al.}(2011)\citenamefont {Radisavljevic}, \citenamefont {Whitwick},\ and\ \citenamefont {Kis}}]{radisavljevic-ACSN-9934}%
  \BibitemOpen
  \bibfield  {author} {\bibinfo {author} {\bibfnamefont {B.}~\bibnamefont {Radisavljevic}}, \bibinfo {author} {\bibfnamefont {M.~B.}\ \bibnamefont {Whitwick}},\ and\ \bibinfo {author} {\bibfnamefont {A.}~\bibnamefont {Kis}},\ }\href {https://doi.org/10.1021/nn203715c} {\bibfield  {journal} {\bibinfo  {journal} {ACS Nano}\ }\textbf {\bibinfo {volume} {5}},\ \bibinfo {pages} {9934} (\bibinfo {year} {2011})}\BibitemShut {NoStop}%
\bibitem [{\citenamefont {Jariwala}\ \emph {et~al.}(2014)\citenamefont {Jariwala}, \citenamefont {Sangwan}, \citenamefont {Lauhon}, \citenamefont {Marks},\ and\ \citenamefont {Hersam}}]{jariwala-ACSN-8-1102}%
  \BibitemOpen
  \bibfield  {author} {\bibinfo {author} {\bibfnamefont {D.}~\bibnamefont {Jariwala}}, \bibinfo {author} {\bibfnamefont {V.~K.}\ \bibnamefont {Sangwan}}, \bibinfo {author} {\bibfnamefont {L.~J.}\ \bibnamefont {Lauhon}}, \bibinfo {author} {\bibfnamefont {T.~J.}\ \bibnamefont {Marks}},\ and\ \bibinfo {author} {\bibfnamefont {M.~C.}\ \bibnamefont {Hersam}},\ }\href {https://doi.org/10.1021/nn500064s} {\bibfield  {journal} {\bibinfo  {journal} {ACS Nano}\ }\textbf {\bibinfo {volume} {8}},\ \bibinfo {pages} {1102} (\bibinfo {year} {2014})}\BibitemShut {NoStop}%
\bibitem [{\citenamefont {{Ryder}}\ \emph {et~al.}(2016)\citenamefont {{Ryder}}, \citenamefont {{Wood}}, \citenamefont {{Wells}},\ and\ \citenamefont {{Hersam}}}]{ryder-ACSN-10-3900}%
  \BibitemOpen
  \bibfield  {author} {\bibinfo {author} {\bibfnamefont {C.~R.}\ \bibnamefont {{Ryder}}}, \bibinfo {author} {\bibfnamefont {J.~D.}\ \bibnamefont {{Wood}}}, \bibinfo {author} {\bibfnamefont {S.~A.}\ \bibnamefont {{Wells}}},\ and\ \bibinfo {author} {\bibfnamefont {M.~C.}\ \bibnamefont {{Hersam}}},\ }\href {https://doi.org/10.1021/acsnano.6b01091} {\bibfield  {journal} {\bibinfo  {journal} {ACS Nano}\ }\textbf {\bibinfo {volume} {10}},\ \bibinfo {pages} {3900–3917} (\bibinfo {year} {2016})}\BibitemShut {NoStop}%
\bibitem [{\citenamefont {{Kam}}\ and\ \citenamefont {{Parkinson}}(1982)}]{kam-JPCY-86-463}%
  \BibitemOpen
  \bibfield  {author} {\bibinfo {author} {\bibfnamefont {K.~K.}\ \bibnamefont {{Kam}}}\ and\ \bibinfo {author} {\bibfnamefont {B.~A.}\ \bibnamefont {{Parkinson}}},\ }\href {https://doi.org/10.1021/j100393a010} {\bibfield  {journal} {\bibinfo  {journal} {J{.} Phys{.} C}\ }\textbf {\bibinfo {volume} {86}},\ \bibinfo {pages} {463–467} (\bibinfo {year} {1982})}\BibitemShut {NoStop}%
\bibitem [{\citenamefont {{Gusakova}}\ \emph {et~al.}(2017)\citenamefont {{Gusakova}}, \citenamefont {{Wang}}, \citenamefont {{Shiau}}, \citenamefont {{Krivosheeva}}, \citenamefont {{Shaposhnikov}}, \citenamefont {{Borisenko}}, \citenamefont {{Gusakov}},\ and\ \citenamefont {{Tay}}}]{gusakova-PSSA-214}%
  \BibitemOpen
  \bibfield  {author} {\bibinfo {author} {\bibfnamefont {J.}~\bibnamefont {{Gusakova}}}, \bibinfo {author} {\bibfnamefont {X.}~\bibnamefont {{Wang}}}, \bibinfo {author} {\bibfnamefont {L.~L.}\ \bibnamefont {{Shiau}}}, \bibinfo {author} {\bibfnamefont {A.}~\bibnamefont {{Krivosheeva}}}, \bibinfo {author} {\bibfnamefont {V.}~\bibnamefont {{Shaposhnikov}}}, \bibinfo {author} {\bibfnamefont {V.}~\bibnamefont {{Borisenko}}}, \bibinfo {author} {\bibfnamefont {V.}~\bibnamefont {{Gusakov}}},\ and\ \bibinfo {author} {\bibfnamefont {B.~K.}\ \bibnamefont {{Tay}}},\ }\href {https://doi.org/10.1002/pssa.201700218} {\bibfield  {journal} {\bibinfo  {journal} {Phys{.} Status Solidi A}\ }\textbf {\bibinfo {volume} {214}},\ \bibinfo {pages} {1700218} (\bibinfo {year} {2017})}\BibitemShut {NoStop}%
\bibitem [{\citenamefont {{Mak}}\ \emph {et~al.}(2010)\citenamefont {{Mak}}, \citenamefont {{Lee}}, \citenamefont {{Hone}}, \citenamefont {{Shan}},\ and\ \citenamefont {{Heinz}}}]{mak-PRL-105}%
  \BibitemOpen
  \bibfield  {author} {\bibinfo {author} {\bibfnamefont {K.~F.}\ \bibnamefont {{Mak}}}, \bibinfo {author} {\bibfnamefont {C.}~\bibnamefont {{Lee}}}, \bibinfo {author} {\bibfnamefont {J.}~\bibnamefont {{Hone}}}, \bibinfo {author} {\bibfnamefont {J.}~\bibnamefont {{Shan}}},\ and\ \bibinfo {author} {\bibfnamefont {T.~F.}\ \bibnamefont {{Heinz}}},\ }\href {https://doi.org/10.1103/physrevlett.105.136805} {\bibfield  {journal} {\bibinfo  {journal} {Phys{.} Rev{.} Lett{.}}\ }\textbf {\bibinfo {volume} {105}},\ \bibinfo {pages} {136805} (\bibinfo {year} {2010})}\BibitemShut {NoStop}%
\bibitem [{\citenamefont {{Island}}\ \emph {et~al.}(2016)\citenamefont {{Island}}, \citenamefont {{Kuc}}, \citenamefont {{Diependaal}}, \citenamefont {{Bratschitsch}}, \citenamefont {{van der Zant}}, \citenamefont {{Heine}},\ and\ \citenamefont {{Castellanos-Gomez}}}]{island-NS-8-2589}%
  \BibitemOpen
  \bibfield  {author} {\bibinfo {author} {\bibfnamefont {J.~O.}\ \bibnamefont {{Island}}}, \bibinfo {author} {\bibfnamefont {A.}~\bibnamefont {{Kuc}}}, \bibinfo {author} {\bibfnamefont {E.~H.}\ \bibnamefont {{Diependaal}}}, \bibinfo {author} {\bibfnamefont {R.}~\bibnamefont {{Bratschitsch}}}, \bibinfo {author} {\bibfnamefont {H.~S.}\ \bibnamefont {{van der Zant}}}, \bibinfo {author} {\bibfnamefont {T.}~\bibnamefont {{Heine}}},\ and\ \bibinfo {author} {\bibfnamefont {A.}~\bibnamefont {{Castellanos-Gomez}}},\ }\href {https://doi.org/10.1039/c5nr08219f} {\bibfield  {journal} {\bibinfo  {journal} {Nanoscale}\ }\textbf {\bibinfo {volume} {8}},\ \bibinfo {pages} {2589–2593} (\bibinfo {year} {2016})}\BibitemShut {NoStop}%
\bibitem [{\citenamefont {{Huang}}\ \emph {et~al.}(2015)\citenamefont {{Huang}}, \citenamefont {{Chen}}, \citenamefont {{Zhang}}, \citenamefont {{Quek}}, \citenamefont {{Chen}}, \citenamefont {{Li}}, \citenamefont {{Hsu}}, \citenamefont {{Chang}}, \citenamefont {{Zheng}}, \citenamefont {{Chen}},\ and\ \citenamefont {{Wee}}}]{huang-NCOMM-6}%
  \BibitemOpen
  \bibfield  {author} {\bibinfo {author} {\bibfnamefont {Y.~L.}\ \bibnamefont {{Huang}}}, \bibinfo {author} {\bibfnamefont {Y.}~\bibnamefont {{Chen}}}, \bibinfo {author} {\bibfnamefont {W.}~\bibnamefont {{Zhang}}}, \bibinfo {author} {\bibfnamefont {S.~Y.}\ \bibnamefont {{Quek}}}, \bibinfo {author} {\bibfnamefont {C.-H.}\ \bibnamefont {{Chen}}}, \bibinfo {author} {\bibfnamefont {L.-J.}\ \bibnamefont {{Li}}}, \bibinfo {author} {\bibfnamefont {W.-T.}\ \bibnamefont {{Hsu}}}, \bibinfo {author} {\bibfnamefont {W.-H.}\ \bibnamefont {{Chang}}}, \bibinfo {author} {\bibfnamefont {Y.~J.}\ \bibnamefont {{Zheng}}}, \bibinfo {author} {\bibfnamefont {W.}~\bibnamefont {{Chen}}},\ and\ \bibinfo {author} {\bibfnamefont {A.~T.~S.}\ \bibnamefont {{Wee}}},\ }\href {https://doi.org/10.1038/ncomms7298} {\bibfield  {journal} {\bibinfo  {journal} {Nature Comm{.}}\ }\textbf {\bibinfo {volume} {6}},\ \bibinfo {pages} {6298} (\bibinfo {year} {2015})}\BibitemShut {NoStop}%
\bibitem [{\citenamefont {{Radisavljevic}}\ \emph {et~al.}(2011)\citenamefont {{Radisavljevic}}, \citenamefont {{Radenovic}}, \citenamefont {{Brivio}}, \citenamefont {{Giacometti}},\ and\ \citenamefont {{Kis}}}]{radisavljevic-NN-6-147}%
  \BibitemOpen
  \bibfield  {author} {\bibinfo {author} {\bibfnamefont {B.}~\bibnamefont {{Radisavljevic}}}, \bibinfo {author} {\bibfnamefont {A.}~\bibnamefont {{Radenovic}}}, \bibinfo {author} {\bibfnamefont {J.}~\bibnamefont {{Brivio}}}, \bibinfo {author} {\bibfnamefont {V.}~\bibnamefont {{Giacometti}}},\ and\ \bibinfo {author} {\bibfnamefont {A.}~\bibnamefont {{Kis}}},\ }\href {https://doi.org/10.1038/nnano.2010.279} {\bibfield  {journal} {\bibinfo  {journal} {Nature Nano{.}}\ }\textbf {\bibinfo {volume} {6}},\ \bibinfo {pages} {147–150} (\bibinfo {year} {2011})}\BibitemShut {NoStop}%
\bibitem [{\citenamefont {{Castellanos‐Gomez}}\ \emph {et~al.}(2012)\citenamefont {{Castellanos‐Gomez}}, \citenamefont {{Poot}}, \citenamefont {{Steele}}, \citenamefont {{van der Zant}}, \citenamefont {{Agraït}},\ and\ \citenamefont {{Rubio‐Bollinger}}}]{castellanos-AM-24-772}%
  \BibitemOpen
  \bibfield  {author} {\bibinfo {author} {\bibfnamefont {A.}~\bibnamefont {{Castellanos‐Gomez}}}, \bibinfo {author} {\bibfnamefont {M.}~\bibnamefont {{Poot}}}, \bibinfo {author} {\bibfnamefont {G.~A.}\ \bibnamefont {{Steele}}}, \bibinfo {author} {\bibfnamefont {H.~S.}\ \bibnamefont {{van der Zant}}}, \bibinfo {author} {\bibfnamefont {N.}~\bibnamefont {{Agraït}}},\ and\ \bibinfo {author} {\bibfnamefont {G.}~\bibnamefont {{Rubio‐Bollinger}}},\ }\href {https://doi.org/10.1002/adma.201103965} {\bibfield  {journal} {\bibinfo  {journal} {Adv{.} Mat{.}}\ }\textbf {\bibinfo {volume} {24}},\ \bibinfo {pages} {772–775} (\bibinfo {year} {2012})}\BibitemShut {NoStop}%
\bibitem [{\citenamefont {{Bernardi}}\ \emph {et~al.}(2013)\citenamefont {{Bernardi}}, \citenamefont {{Palummo}},\ and\ \citenamefont {{Grossman}}}]{bernardi-NL-13-3664}%
  \BibitemOpen
  \bibfield  {author} {\bibinfo {author} {\bibfnamefont {M.}~\bibnamefont {{Bernardi}}}, \bibinfo {author} {\bibfnamefont {M.}~\bibnamefont {{Palummo}}},\ and\ \bibinfo {author} {\bibfnamefont {J.~C.}\ \bibnamefont {{Grossman}}},\ }\href {https://doi.org/10.1021/nl401544y} {\bibfield  {journal} {\bibinfo  {journal} {Nano Letters}\ }\textbf {\bibinfo {volume} {13}},\ \bibinfo {pages} {3664–3670} (\bibinfo {year} {2013})}\BibitemShut {NoStop}%
\bibitem [{\citenamefont {{Jariwala}}\ \emph {et~al.}(2017)\citenamefont {{Jariwala}}, \citenamefont {{Davoyan}}, \citenamefont {{Wong}},\ and\ \citenamefont {{Atwater}}}]{jariwala-ACSP-12-2962}%
  \BibitemOpen
  \bibfield  {author} {\bibinfo {author} {\bibfnamefont {D.}~\bibnamefont {{Jariwala}}}, \bibinfo {author} {\bibfnamefont {A.~R.}\ \bibnamefont {{Davoyan}}}, \bibinfo {author} {\bibfnamefont {J.}~\bibnamefont {{Wong}}},\ and\ \bibinfo {author} {\bibfnamefont {H.~A.}\ \bibnamefont {{Atwater}}},\ }\href {https://doi.org/10.1021/acsphotonics.7b01103} {\bibfield  {journal} {\bibinfo  {journal} {ACS Photonics}\ }\textbf {\bibinfo {volume} {4}},\ \bibinfo {pages} {2962–2970} (\bibinfo {year} {2017})}\BibitemShut {NoStop}%
\bibitem [{\citenamefont {{Kim}}\ \emph {et~al.}(2022)\citenamefont {{Kim}}, \citenamefont {{Andreev}}, \citenamefont {{Choi}}, \citenamefont {{Shim}}, \citenamefont {{Ahn}}, \citenamefont {{Lynch}}, \citenamefont {{Lee}}, \citenamefont {{Lee}}, \citenamefont {{Nazif}}, \citenamefont {{Kumar}}, \citenamefont {{Kumar}}, \citenamefont {{Choo}}, \citenamefont {{Jariwala}}, \citenamefont {{Saraswat}},\ and\ \citenamefont {{Park}}}]{kim-ACSN-16-8827}%
  \BibitemOpen
  \bibfield  {author} {\bibinfo {author} {\bibfnamefont {K.-H.}\ \bibnamefont {{Kim}}}, \bibinfo {author} {\bibfnamefont {M.}~\bibnamefont {{Andreev}}}, \bibinfo {author} {\bibfnamefont {S.}~\bibnamefont {{Choi}}}, \bibinfo {author} {\bibfnamefont {J.}~\bibnamefont {{Shim}}}, \bibinfo {author} {\bibfnamefont {H.}~\bibnamefont {{Ahn}}}, \bibinfo {author} {\bibfnamefont {J.}~\bibnamefont {{Lynch}}}, \bibinfo {author} {\bibfnamefont {T.}~\bibnamefont {{Lee}}}, \bibinfo {author} {\bibfnamefont {J.}~\bibnamefont {{Lee}}}, \bibinfo {author} {\bibfnamefont {K.~N.}\ \bibnamefont {{Nazif}}}, \bibinfo {author} {\bibfnamefont {A.}~\bibnamefont {{Kumar}}}, \bibinfo {author} {\bibfnamefont {P.}~\bibnamefont {{Kumar}}}, \bibinfo {author} {\bibfnamefont {H.}~\bibnamefont {{Choo}}}, \bibinfo {author} {\bibfnamefont {D.}~\bibnamefont {{Jariwala}}}, \bibinfo {author} {\bibfnamefont {K.~C.}\ \bibnamefont {{Saraswat}}},\ and\ \bibinfo {author} {\bibfnamefont {J.-H.}\ \bibnamefont {{Park}}},\ }\href
  {https://doi.org/10.1021/acsnano.1c10054} {\bibfield  {journal} {\bibinfo  {journal} {ACS Nano}\ }\textbf {\bibinfo {volume} {16}},\ \bibinfo {pages} {8827–8836} (\bibinfo {year} {2022})}\BibitemShut {NoStop}%
\bibitem [{\citenamefont {{Tsai}}\ \emph {et~al.}(2017)\citenamefont {{Tsai}}, \citenamefont {{Li}}, \citenamefont {{Retamal}}, \citenamefont {{Lam}}, \citenamefont {{Lin}}, \citenamefont {{Suenaga}}, \citenamefont {{Chen}}, \citenamefont {{Liang}}, \citenamefont {{Li}},\ and\ \citenamefont {{He}}}]{tsai-AM-29}%
  \BibitemOpen
  \bibfield  {author} {\bibinfo {author} {\bibfnamefont {M.}~\bibnamefont {{Tsai}}}, \bibinfo {author} {\bibfnamefont {M.}~\bibnamefont {{Li}}}, \bibinfo {author} {\bibfnamefont {J.~R.}\ \bibnamefont {{Retamal}}}, \bibinfo {author} {\bibfnamefont {K.}~\bibnamefont {{Lam}}}, \bibinfo {author} {\bibfnamefont {Y.}~\bibnamefont {{Lin}}}, \bibinfo {author} {\bibfnamefont {K.}~\bibnamefont {{Suenaga}}}, \bibinfo {author} {\bibfnamefont {L.}~\bibnamefont {{Chen}}}, \bibinfo {author} {\bibfnamefont {G.}~\bibnamefont {{Liang}}}, \bibinfo {author} {\bibfnamefont {L.}~\bibnamefont {{Li}}},\ and\ \bibinfo {author} {\bibfnamefont {J.}~\bibnamefont {{He}}},\ }\href {https://doi.org/10.1002/adma.201701168} {\bibfield  {journal} {\bibinfo  {journal} {Adv{.} Mat{.}}\ }\textbf {\bibinfo {volume} {29}},\ \bibinfo {pages} {1701168} (\bibinfo {year} {2017})}\BibitemShut {NoStop}%
\bibitem [{\citenamefont {{Kwak}}\ \emph {et~al.}(2018)\citenamefont {{Kwak}}, \citenamefont {{Ra}}, \citenamefont {{Jeong}}, \citenamefont {{Lee}},\ and\ \citenamefont {{Lee}}}]{kwak-ADVMI-5}%
  \BibitemOpen
  \bibfield  {author} {\bibinfo {author} {\bibfnamefont {D.}~\bibnamefont {{Kwak}}}, \bibinfo {author} {\bibfnamefont {H.}~\bibnamefont {{Ra}}}, \bibinfo {author} {\bibfnamefont {M.}~\bibnamefont {{Jeong}}}, \bibinfo {author} {\bibfnamefont {A.}~\bibnamefont {{Lee}}},\ and\ \bibinfo {author} {\bibfnamefont {J.}~\bibnamefont {{Lee}}},\ }\href {https://doi.org/10.1002/admi.201800671} {\bibfield  {journal} {\bibinfo  {journal} {Adv{.} Mat{.} Interfaces}\ }\textbf {\bibinfo {volume} {5}},\ \bibinfo {pages} {1800671} (\bibinfo {year} {2018})}\BibitemShut {NoStop}%
\bibitem [{\citenamefont {{Nassiri Nazif}}\ \emph {et~al.}(2021)\citenamefont {{Nassiri Nazif}}, \citenamefont {{Daus}}, \citenamefont {{Hong}}, \citenamefont {{Lee}}, \citenamefont {{Vaziri}}, \citenamefont {{Kumar}}, \citenamefont {{Nitta}}, \citenamefont {{Chen}}, \citenamefont {{Kananian}}, \citenamefont {{Islam}}, \citenamefont {{Kim}}, \citenamefont {{Park}}, \citenamefont {{Poon}}, \citenamefont {{Brongersma}}, \citenamefont {{Pop}},\ and\ \citenamefont {{Saraswat}}}]{nassiri-NCOMM-12}%
  \BibitemOpen
  \bibfield  {author} {\bibinfo {author} {\bibfnamefont {K.}~\bibnamefont {{Nassiri Nazif}}}, \bibinfo {author} {\bibfnamefont {A.}~\bibnamefont {{Daus}}}, \bibinfo {author} {\bibfnamefont {J.}~\bibnamefont {{Hong}}}, \bibinfo {author} {\bibfnamefont {N.}~\bibnamefont {{Lee}}}, \bibinfo {author} {\bibfnamefont {S.}~\bibnamefont {{Vaziri}}}, \bibinfo {author} {\bibfnamefont {A.}~\bibnamefont {{Kumar}}}, \bibinfo {author} {\bibfnamefont {F.}~\bibnamefont {{Nitta}}}, \bibinfo {author} {\bibfnamefont {M.~E.}\ \bibnamefont {{Chen}}}, \bibinfo {author} {\bibfnamefont {S.}~\bibnamefont {{Kananian}}}, \bibinfo {author} {\bibfnamefont {R.}~\bibnamefont {{Islam}}}, \bibinfo {author} {\bibfnamefont {K.-H.}\ \bibnamefont {{Kim}}}, \bibinfo {author} {\bibfnamefont {J.-H.}\ \bibnamefont {{Park}}}, \bibinfo {author} {\bibfnamefont {A.~S.~Y.}\ \bibnamefont {{Poon}}}, \bibinfo {author} {\bibfnamefont {M.~L.}\ \bibnamefont {{Brongersma}}}, \bibinfo {author} {\bibfnamefont {E.}~\bibnamefont {{Pop}}},\ and\ \bibinfo {author}
  {\bibfnamefont {K.~C.}\ \bibnamefont {{Saraswat}}},\ }\href {https://doi.org/10.1038/s41467-021-27195-7} {\bibfield  {journal} {\bibinfo  {journal} {Nature Comm{.}}\ }\textbf {\bibinfo {volume} {12}},\ \bibinfo {pages} {7034} (\bibinfo {year} {2021})}\BibitemShut {NoStop}%
\bibitem [{\citenamefont {Tongay}\ \emph {et~al.}(2013)\citenamefont {Tongay}, \citenamefont {Suh}, \citenamefont {Ataca}, \citenamefont {Fan}, \citenamefont {Luce}, \citenamefont {Kang}, \citenamefont {Liu}, \citenamefont {Ko}, \citenamefont {Raghunathanan}, \citenamefont {Zhou}, \citenamefont {Ogletree}, \citenamefont {Li}, \citenamefont {Grossman},\ and\ \citenamefont {Wu}}]{tongay-SR-3}%
  \BibitemOpen
  \bibfield  {author} {\bibinfo {author} {\bibfnamefont {S.}~\bibnamefont {Tongay}}, \bibinfo {author} {\bibfnamefont {J.}~\bibnamefont {Suh}}, \bibinfo {author} {\bibfnamefont {C.}~\bibnamefont {Ataca}}, \bibinfo {author} {\bibfnamefont {W.}~\bibnamefont {Fan}}, \bibinfo {author} {\bibfnamefont {A.}~\bibnamefont {Luce}}, \bibinfo {author} {\bibfnamefont {J.~S.}\ \bibnamefont {Kang}}, \bibinfo {author} {\bibfnamefont {J.}~\bibnamefont {Liu}}, \bibinfo {author} {\bibfnamefont {C.}~\bibnamefont {Ko}}, \bibinfo {author} {\bibfnamefont {R.}~\bibnamefont {Raghunathanan}}, \bibinfo {author} {\bibfnamefont {J.}~\bibnamefont {Zhou}}, \bibinfo {author} {\bibfnamefont {F.}~\bibnamefont {Ogletree}}, \bibinfo {author} {\bibfnamefont {J.}~\bibnamefont {Li}}, \bibinfo {author} {\bibfnamefont {J.~C.}\ \bibnamefont {Grossman}},\ and\ \bibinfo {author} {\bibfnamefont {J.}~\bibnamefont {Wu}},\ }\href {https://doi.org/10.1038/srep02657} {\bibfield  {journal} {\bibinfo  {journal} {Scientific Reports}\ }\textbf {\bibinfo {volume}
  {3}},\ \bibinfo {pages} {2657} (\bibinfo {year} {2013})}\BibitemShut {NoStop}%
\bibitem [{\citenamefont {Nassiri~Nazif}\ \emph {et~al.}(2021)\citenamefont {Nassiri~Nazif}, \citenamefont {Kumar}, \citenamefont {Hong}, \citenamefont {Lee}, \citenamefont {Islam}, \citenamefont {McClellan}, \citenamefont {Karni}, \citenamefont {van~de Groep}, \citenamefont {Heinz}, \citenamefont {Pop}, \citenamefont {Brongersma},\ and\ \citenamefont {Saraswat}}]{nassiri-NL-21-3443}%
  \BibitemOpen
  \bibfield  {author} {\bibinfo {author} {\bibfnamefont {K.}~\bibnamefont {Nassiri~Nazif}}, \bibinfo {author} {\bibfnamefont {A.}~\bibnamefont {Kumar}}, \bibinfo {author} {\bibfnamefont {J.}~\bibnamefont {Hong}}, \bibinfo {author} {\bibfnamefont {N.}~\bibnamefont {Lee}}, \bibinfo {author} {\bibfnamefont {R.}~\bibnamefont {Islam}}, \bibinfo {author} {\bibfnamefont {C.~J.}\ \bibnamefont {McClellan}}, \bibinfo {author} {\bibfnamefont {O.}~\bibnamefont {Karni}}, \bibinfo {author} {\bibfnamefont {J.}~\bibnamefont {van~de Groep}}, \bibinfo {author} {\bibfnamefont {T.~F.}\ \bibnamefont {Heinz}}, \bibinfo {author} {\bibfnamefont {E.}~\bibnamefont {Pop}}, \bibinfo {author} {\bibfnamefont {M.~L.}\ \bibnamefont {Brongersma}},\ and\ \bibinfo {author} {\bibfnamefont {K.~C.}\ \bibnamefont {Saraswat}},\ }\href {https://doi.org/10.1021/acs.nanolett.1c00015} {\bibfield  {journal} {\bibinfo  {journal} {Nano Letters}\ }\textbf {\bibinfo {volume} {21}},\ \bibinfo {pages} {3443–3450} (\bibinfo {year} {2021})}\BibitemShut
  {NoStop}%
\bibitem [{\citenamefont {Zhou}\ \emph {et~al.}(2013)\citenamefont {Zhou}, \citenamefont {Zou}, \citenamefont {Najmaei}, \citenamefont {Liu}, \citenamefont {Shi}, \citenamefont {Kong}, \citenamefont {Lou}, \citenamefont {Ajayan}, \citenamefont {Yakobson},\ and\ \citenamefont {Idrobo}}]{zhou-NL-13-2615}%
  \BibitemOpen
  \bibfield  {author} {\bibinfo {author} {\bibfnamefont {W.}~\bibnamefont {Zhou}}, \bibinfo {author} {\bibfnamefont {X.}~\bibnamefont {Zou}}, \bibinfo {author} {\bibfnamefont {S.}~\bibnamefont {Najmaei}}, \bibinfo {author} {\bibfnamefont {Z.}~\bibnamefont {Liu}}, \bibinfo {author} {\bibfnamefont {Y.}~\bibnamefont {Shi}}, \bibinfo {author} {\bibfnamefont {J.}~\bibnamefont {Kong}}, \bibinfo {author} {\bibfnamefont {J.}~\bibnamefont {Lou}}, \bibinfo {author} {\bibfnamefont {P.~M.}\ \bibnamefont {Ajayan}}, \bibinfo {author} {\bibfnamefont {B.~I.}\ \bibnamefont {Yakobson}},\ and\ \bibinfo {author} {\bibfnamefont {J.-C.}\ \bibnamefont {Idrobo}},\ }\href {https://doi.org/10.1021/nl4007479} {\bibfield  {journal} {\bibinfo  {journal} {Nano Letters}\ }\textbf {\bibinfo {volume} {13}},\ \bibinfo {pages} {2615} (\bibinfo {year} {2013})}\BibitemShut {NoStop}%
\bibitem [{\citenamefont {Xu}\ \emph {et~al.}(2020)\citenamefont {Xu}, \citenamefont {Yang},\ and\ \citenamefont {Zheng}}]{xu-MTC-22-100772}%
  \BibitemOpen
  \bibfield  {author} {\bibinfo {author} {\bibfnamefont {Q.}~\bibnamefont {Xu}}, \bibinfo {author} {\bibfnamefont {G.}~\bibnamefont {Yang}},\ and\ \bibinfo {author} {\bibfnamefont {W.}~\bibnamefont {Zheng}},\ }\href {https://doi.org/10.1016/j.mtcomm.2019.100772} {\bibfield  {journal} {\bibinfo  {journal} {Mater{.} Today Comm{.}}\ }\textbf {\bibinfo {volume} {22}},\ \bibinfo {pages} {100772} (\bibinfo {year} {2020})}\BibitemShut {NoStop}%
\bibitem [{\citenamefont {Tan}\ \emph {et~al.}(2020)\citenamefont {Tan}, \citenamefont {Freysoldt},\ and\ \citenamefont {Hennig}}]{tan-PRM-4-064004}%
  \BibitemOpen
  \bibfield  {author} {\bibinfo {author} {\bibfnamefont {A.~M.~Z.}\ \bibnamefont {Tan}}, \bibinfo {author} {\bibfnamefont {C.}~\bibnamefont {Freysoldt}},\ and\ \bibinfo {author} {\bibfnamefont {R.~G.}\ \bibnamefont {Hennig}},\ }\href {https://doi.org/10.1103/PhysRevMaterials.4.064004} {\ \textbf {\bibinfo {volume} {4}},\ \bibinfo {pages} {064004} (\bibinfo {year} {2020})}\BibitemShut {NoStop}%
\bibitem [{\citenamefont {Haldar}\ \emph {et~al.}(2015)\citenamefont {Haldar}, \citenamefont {Vovusha}, \citenamefont {Yadav}, \citenamefont {Eriksson},\ and\ \citenamefont {Sanyal}}]{haldar-PRB-92-235408}%
  \BibitemOpen
  \bibfield  {author} {\bibinfo {author} {\bibfnamefont {S.}~\bibnamefont {Haldar}}, \bibinfo {author} {\bibfnamefont {H.}~\bibnamefont {Vovusha}}, \bibinfo {author} {\bibfnamefont {M.~K.}\ \bibnamefont {Yadav}}, \bibinfo {author} {\bibfnamefont {O.}~\bibnamefont {Eriksson}},\ and\ \bibinfo {author} {\bibfnamefont {B.}~\bibnamefont {Sanyal}},\ }\href {https://doi.org/10.1103/physrevb.92.235408} {\bibfield  {journal} {\bibinfo  {journal} {Phys{.} Rev{.} B}\ }\textbf {\bibinfo {volume} {92}},\ \bibinfo {pages} {235408} (\bibinfo {year} {2015})}\BibitemShut {NoStop}%
\bibitem [{\citenamefont {Komsa}\ and\ \citenamefont {Krasheninnikov}(2015)}]{komsa-PRB-91-125304}%
  \BibitemOpen
  \bibfield  {author} {\bibinfo {author} {\bibfnamefont {H.-P.}\ \bibnamefont {Komsa}}\ and\ \bibinfo {author} {\bibfnamefont {A.~V.}\ \bibnamefont {Krasheninnikov}},\ }\href {https://doi.org/10.1103/physrevb.91.125304} {\bibfield  {journal} {\bibinfo  {journal} {Phys{.} Rev{.} B}\ }\textbf {\bibinfo {volume} {91}},\ \bibinfo {pages} {125304} (\bibinfo {year} {2015})}\BibitemShut {NoStop}%
\bibitem [{\citenamefont {Noh}\ \emph {et~al.}(2014)\citenamefont {Noh}, \citenamefont {Kim},\ and\ \citenamefont {Kim}}]{noh-PRB-89-205417}%
  \BibitemOpen
  \bibfield  {author} {\bibinfo {author} {\bibfnamefont {J.-Y.}\ \bibnamefont {Noh}}, \bibinfo {author} {\bibfnamefont {H.}~\bibnamefont {Kim}},\ and\ \bibinfo {author} {\bibfnamefont {Y.-S.}\ \bibnamefont {Kim}},\ }\href {https://doi.org/10.1103/PhysRevB.89.205417} {\bibfield  {journal} {\bibinfo  {journal} {Phys{.} Rev{.} B}\ }\textbf {\bibinfo {volume} {89}},\ \bibinfo {pages} {205417} (\bibinfo {year} {2014})}\BibitemShut {NoStop}%
\bibitem [{\citenamefont {Gusakov}\ \emph {et~al.}(2021)\citenamefont {Gusakov}, \citenamefont {Gusakova},\ and\ \citenamefont {Tay}}]{gusakov-PSSB-259-2100479}%
  \BibitemOpen
  \bibfield  {author} {\bibinfo {author} {\bibfnamefont {V.}~\bibnamefont {Gusakov}}, \bibinfo {author} {\bibfnamefont {J.}~\bibnamefont {Gusakova}},\ and\ \bibinfo {author} {\bibfnamefont {B.~K.}\ \bibnamefont {Tay}},\ }\href {https://doi.org/10.1002/pssb.202100479} {\bibfield  {journal} {\bibinfo  {journal} {Phys{.} Status Solidi B}\ }\textbf {\bibinfo {volume} {259}},\ \bibinfo {pages} {2100479} (\bibinfo {year} {2021})}\BibitemShut {NoStop}%
\bibitem [{\citenamefont {KC}\ \emph {et~al.}(2014)\citenamefont {KC}, \citenamefont {Longo}, \citenamefont {Addou}, \citenamefont {Wallace},\ and\ \citenamefont {Cho}}]{santosh-NA-375703}%
  \BibitemOpen
  \bibfield  {author} {\bibinfo {author} {\bibfnamefont {S.}~\bibnamefont {KC}}, \bibinfo {author} {\bibfnamefont {R.~C.}\ \bibnamefont {Longo}}, \bibinfo {author} {\bibfnamefont {R.}~\bibnamefont {Addou}}, \bibinfo {author} {\bibfnamefont {R.~M.}\ \bibnamefont {Wallace}},\ and\ \bibinfo {author} {\bibfnamefont {K.}~\bibnamefont {Cho}},\ }\href {https://doi.org/10.1088/0957-4484/25/37/375703} {\bibfield  {journal} {\bibinfo  {journal} {Nanotechnol{.}}\ }\textbf {\bibinfo {volume} {25}},\ \bibinfo {pages} {375703} (\bibinfo {year} {2014})}\BibitemShut {NoStop}%
\bibitem [{\citenamefont {Liang}\ \emph {et~al.}(2020)\citenamefont {Liang}, \citenamefont {Habib}, \citenamefont {Xiao}, \citenamefont {Xie}, \citenamefont {Kong}, \citenamefont {Yu}, \citenamefont {Iwai}, \citenamefont {Fujita}, \citenamefont {Hanagata}, \citenamefont {Chen}, \citenamefont {Feng},\ and\ \citenamefont {Xu}}]{liang-ACSAEM-2-1055}%
  \BibitemOpen
  \bibfield  {author} {\bibinfo {author} {\bibfnamefont {T.}~\bibnamefont {Liang}}, \bibinfo {author} {\bibfnamefont {M.~R.}\ \bibnamefont {Habib}}, \bibinfo {author} {\bibfnamefont {H.}~\bibnamefont {Xiao}}, \bibinfo {author} {\bibfnamefont {S.}~\bibnamefont {Xie}}, \bibinfo {author} {\bibfnamefont {Y.}~\bibnamefont {Kong}}, \bibinfo {author} {\bibfnamefont {C.}~\bibnamefont {Yu}}, \bibinfo {author} {\bibfnamefont {H.}~\bibnamefont {Iwai}}, \bibinfo {author} {\bibfnamefont {D.}~\bibnamefont {Fujita}}, \bibinfo {author} {\bibfnamefont {N.}~\bibnamefont {Hanagata}}, \bibinfo {author} {\bibfnamefont {H.}~\bibnamefont {Chen}}, \bibinfo {author} {\bibfnamefont {Z.}~\bibnamefont {Feng}},\ and\ \bibinfo {author} {\bibfnamefont {M.}~\bibnamefont {Xu}},\ }\href {https://doi.org/10.1021/acsaelm.0c00076} {\bibfield  {journal} {\bibinfo  {journal} {ACS Appl{.} Electronic Mat{.}}\ }\textbf {\bibinfo {volume} {2}},\ \bibinfo {pages} {1055} (\bibinfo {year} {2020})}\BibitemShut {NoStop}%
\bibitem [{\citenamefont {Schaefer}\ \emph {et~al.}(2021)\citenamefont {Schaefer}, \citenamefont {Caicedo~Roque}, \citenamefont {Sauthier}, \citenamefont {Bousquet}, \citenamefont {Hébert}, \citenamefont {Sperling}, \citenamefont {Pérez-Tomás}, \citenamefont {Santiso}, \citenamefont {del Corro},\ and\ \citenamefont {Garrido}}]{schaefer-CM-4474}%
  \BibitemOpen
  \bibfield  {author} {\bibinfo {author} {\bibfnamefont {C.~M.}\ \bibnamefont {Schaefer}}, \bibinfo {author} {\bibfnamefont {J.~M.}\ \bibnamefont {Caicedo~Roque}}, \bibinfo {author} {\bibfnamefont {G.}~\bibnamefont {Sauthier}}, \bibinfo {author} {\bibfnamefont {J.}~\bibnamefont {Bousquet}}, \bibinfo {author} {\bibfnamefont {C.}~\bibnamefont {Hébert}}, \bibinfo {author} {\bibfnamefont {J.~R.}\ \bibnamefont {Sperling}}, \bibinfo {author} {\bibfnamefont {A.}~\bibnamefont {Pérez-Tomás}}, \bibinfo {author} {\bibfnamefont {J.}~\bibnamefont {Santiso}}, \bibinfo {author} {\bibfnamefont {E.}~\bibnamefont {del Corro}},\ and\ \bibinfo {author} {\bibfnamefont {J.~A.}\ \bibnamefont {Garrido}},\ }\href {https://doi.org/10.1021/acs.chemmater.1c00646} {\bibfield  {journal} {\bibinfo  {journal} {Chem{.} of Mat{.}}\ }\textbf {\bibinfo {volume} {33}},\ \bibinfo {pages} {4474–4487} (\bibinfo {year} {2021})}\BibitemShut {NoStop}%
\bibitem [{\citenamefont {{Ghiami}}\ \emph {et~al.}(2023)\citenamefont {{Ghiami}}, \citenamefont {{Grundmann}}, \citenamefont {{Tang}}, \citenamefont {{Fiadziushkin}}, \citenamefont {{Wang}}, \citenamefont {{Aussen}}, \citenamefont {{Hoffmann-Eifert}}, \citenamefont {{Heuken}}, \citenamefont {{Kalisch}},\ and\ \citenamefont {{Vescan}}}]{ghiami-SUR-6-351}%
  \BibitemOpen
  \bibfield  {author} {\bibinfo {author} {\bibfnamefont {A.}~\bibnamefont {{Ghiami}}}, \bibinfo {author} {\bibfnamefont {A.}~\bibnamefont {{Grundmann}}}, \bibinfo {author} {\bibfnamefont {S.}~\bibnamefont {{Tang}}}, \bibinfo {author} {\bibfnamefont {H.}~\bibnamefont {{Fiadziushkin}}}, \bibinfo {author} {\bibfnamefont {Z.}~\bibnamefont {{Wang}}}, \bibinfo {author} {\bibfnamefont {S.}~\bibnamefont {{Aussen}}}, \bibinfo {author} {\bibfnamefont {S.}~\bibnamefont {{Hoffmann-Eifert}}}, \bibinfo {author} {\bibfnamefont {M.}~\bibnamefont {{Heuken}}}, \bibinfo {author} {\bibfnamefont {H.}~\bibnamefont {{Kalisch}}},\ and\ \bibinfo {author} {\bibfnamefont {A.}~\bibnamefont {{Vescan}}},\ }\href {https://doi.org/10.3390/surfaces6040025} {\bibfield  {journal} {\bibinfo  {journal} {Surfaces}\ }\textbf {\bibinfo {volume} {6}},\ \bibinfo {pages} {351–363} (\bibinfo {year} {2023})}\BibitemShut {NoStop}%
\bibitem [{\citenamefont {{Park}}\ \emph {et~al.}(2023)\citenamefont {{Park}}, \citenamefont {{Li}}, \citenamefont {{Jung}}, \citenamefont {{Singh}}, \citenamefont {{Baik}}, \citenamefont {{Lee}}, \citenamefont {{Oh}}, \citenamefont {{Kim}}, \citenamefont {{Lee}}, \citenamefont {{Woo}}, \citenamefont {{Park}}, \citenamefont {{Kim}}, \citenamefont {{Lee}}, \citenamefont {{Lee}},\ and\ \citenamefont {{Hwang}}}]{park-NPJ2DMA-7-60}%
  \BibitemOpen
  \bibfield  {author} {\bibinfo {author} {\bibfnamefont {Y.}~\bibnamefont {{Park}}}, \bibinfo {author} {\bibfnamefont {N.}~\bibnamefont {{Li}}}, \bibinfo {author} {\bibfnamefont {D.}~\bibnamefont {{Jung}}}, \bibinfo {author} {\bibfnamefont {L.~T.}\ \bibnamefont {{Singh}}}, \bibinfo {author} {\bibfnamefont {J.}~\bibnamefont {{Baik}}}, \bibinfo {author} {\bibfnamefont {E.}~\bibnamefont {{Lee}}}, \bibinfo {author} {\bibfnamefont {D.}~\bibnamefont {{Oh}}}, \bibinfo {author} {\bibfnamefont {Y.~D.}\ \bibnamefont {{Kim}}}, \bibinfo {author} {\bibfnamefont {J.~Y.}\ \bibnamefont {{Lee}}}, \bibinfo {author} {\bibfnamefont {J.}~\bibnamefont {{Woo}}}, \bibinfo {author} {\bibfnamefont {S.}~\bibnamefont {{Park}}}, \bibinfo {author} {\bibfnamefont {H.}~\bibnamefont {{Kim}}}, \bibinfo {author} {\bibfnamefont {G.}~\bibnamefont {{Lee}}}, \bibinfo {author} {\bibfnamefont {G.}~\bibnamefont {{Lee}}},\ and\ \bibinfo {author} {\bibfnamefont {C.-C.}\ \bibnamefont {{Hwang}}},\ }\href {https://doi.org/10.1038/s41699-023-00424-x}
  {\bibfield  {journal} {\bibinfo  {journal} {npj 2D Mat{.} and Appl{.}}\ }\textbf {\bibinfo {volume} {7}},\ \bibinfo {pages} {60} (\bibinfo {year} {2023})}\BibitemShut {NoStop}%
\bibitem [{\citenamefont {Kim}\ \emph {et~al.}(2012)\citenamefont {Kim}, \citenamefont {Konar}, \citenamefont {Hwang}, \citenamefont {Lee}, \citenamefont {Lee}, \citenamefont {Yang}, \citenamefont {Jung}, \citenamefont {Kim}, \citenamefont {Yoo}, \citenamefont {Choi}, \citenamefont {Jin}, \citenamefont {Lee}, \citenamefont {Jena}, \citenamefont {Choi},\ and\ \citenamefont {Kim}}]{kim-NCOMM-1038}%
  \BibitemOpen
  \bibfield  {author} {\bibinfo {author} {\bibfnamefont {S.}~\bibnamefont {Kim}}, \bibinfo {author} {\bibfnamefont {A.}~\bibnamefont {Konar}}, \bibinfo {author} {\bibfnamefont {W.-S.}\ \bibnamefont {Hwang}}, \bibinfo {author} {\bibfnamefont {J.~H.}\ \bibnamefont {Lee}}, \bibinfo {author} {\bibfnamefont {J.}~\bibnamefont {Lee}}, \bibinfo {author} {\bibfnamefont {J.}~\bibnamefont {Yang}}, \bibinfo {author} {\bibfnamefont {C.}~\bibnamefont {Jung}}, \bibinfo {author} {\bibfnamefont {H.}~\bibnamefont {Kim}}, \bibinfo {author} {\bibfnamefont {J.-B.}\ \bibnamefont {Yoo}}, \bibinfo {author} {\bibfnamefont {J.-Y.}\ \bibnamefont {Choi}}, \bibinfo {author} {\bibfnamefont {Y.~W.}\ \bibnamefont {Jin}}, \bibinfo {author} {\bibfnamefont {S.~Y.}\ \bibnamefont {Lee}}, \bibinfo {author} {\bibfnamefont {D.}~\bibnamefont {Jena}}, \bibinfo {author} {\bibfnamefont {W.}~\bibnamefont {Choi}},\ and\ \bibinfo {author} {\bibfnamefont {K.}~\bibnamefont {Kim}},\ }\href {https://doi.org/10.1038/ncomms2018} {\bibfield  {journal} {\bibinfo
  {journal} {Nature Comm{.}}\ }\textbf {\bibinfo {volume} {3}},\ \bibinfo {pages} {1011} (\bibinfo {year} {2012})}\BibitemShut {NoStop}%
\bibitem [{\citenamefont {Yin}\ \emph {et~al.}(2011)\citenamefont {Yin}, \citenamefont {Li}, \citenamefont {Li}, \citenamefont {Jiang}, \citenamefont {Shi}, \citenamefont {Sun}, \citenamefont {Lu}, \citenamefont {Zhang}, \citenamefont {Chen},\ and\ \citenamefont {Zhang}}]{yin-ACSN-74}%
  \BibitemOpen
  \bibfield  {author} {\bibinfo {author} {\bibfnamefont {Z.}~\bibnamefont {Yin}}, \bibinfo {author} {\bibfnamefont {H.}~\bibnamefont {Li}}, \bibinfo {author} {\bibfnamefont {H.}~\bibnamefont {Li}}, \bibinfo {author} {\bibfnamefont {L.}~\bibnamefont {Jiang}}, \bibinfo {author} {\bibfnamefont {Y.}~\bibnamefont {Shi}}, \bibinfo {author} {\bibfnamefont {Y.}~\bibnamefont {Sun}}, \bibinfo {author} {\bibfnamefont {G.}~\bibnamefont {Lu}}, \bibinfo {author} {\bibfnamefont {Q.}~\bibnamefont {Zhang}}, \bibinfo {author} {\bibfnamefont {X.}~\bibnamefont {Chen}},\ and\ \bibinfo {author} {\bibfnamefont {H.}~\bibnamefont {Zhang}},\ }\href {https://doi.org/10.1021/nn2024557} {\bibfield  {journal} {\bibinfo  {journal} {ACS Nano}\ }\textbf {\bibinfo {volume} {6}},\ \bibinfo {pages} {74–80} (\bibinfo {year} {2011})}\BibitemShut {NoStop}%
\bibitem [{\citenamefont {Schmidt}\ \emph {et~al.}(2014)\citenamefont {Schmidt}, \citenamefont {Wang}, \citenamefont {Chu}, \citenamefont {Toh}, \citenamefont {Kumar}, \citenamefont {Zhao}, \citenamefont {Castro~Neto}, \citenamefont {Martin}, \citenamefont {Adam}, \citenamefont {Özyilmaz},\ and\ \citenamefont {Eda}}]{schmidt-NL-1909}%
  \BibitemOpen
  \bibfield  {author} {\bibinfo {author} {\bibfnamefont {H.}~\bibnamefont {Schmidt}}, \bibinfo {author} {\bibfnamefont {S.}~\bibnamefont {Wang}}, \bibinfo {author} {\bibfnamefont {L.}~\bibnamefont {Chu}}, \bibinfo {author} {\bibfnamefont {M.}~\bibnamefont {Toh}}, \bibinfo {author} {\bibfnamefont {R.}~\bibnamefont {Kumar}}, \bibinfo {author} {\bibfnamefont {W.}~\bibnamefont {Zhao}}, \bibinfo {author} {\bibfnamefont {A.~H.}\ \bibnamefont {Castro~Neto}}, \bibinfo {author} {\bibfnamefont {J.}~\bibnamefont {Martin}}, \bibinfo {author} {\bibfnamefont {S.}~\bibnamefont {Adam}}, \bibinfo {author} {\bibfnamefont {B.}~\bibnamefont {Özyilmaz}},\ and\ \bibinfo {author} {\bibfnamefont {G.}~\bibnamefont {Eda}},\ }\href {https://doi.org/10.1021/nl4046922} {\bibfield  {journal} {\bibinfo  {journal} {Nano Letters}\ }\textbf {\bibinfo {volume} {14}},\ \bibinfo {pages} {1909–1913} (\bibinfo {year} {2014})}\BibitemShut {NoStop}%
\bibitem [{\citenamefont {Li}\ \emph {et~al.}(2011)\citenamefont {Li}, \citenamefont {Yin}, \citenamefont {He}, \citenamefont {Li}, \citenamefont {Huang}, \citenamefont {Lu}, \citenamefont {Fam}, \citenamefont {Tok}, \citenamefont {Zhang},\ and\ \citenamefont {Zhang}}]{li-SML-63}%
  \BibitemOpen
  \bibfield  {author} {\bibinfo {author} {\bibfnamefont {H.}~\bibnamefont {Li}}, \bibinfo {author} {\bibfnamefont {Z.}~\bibnamefont {Yin}}, \bibinfo {author} {\bibfnamefont {Q.}~\bibnamefont {He}}, \bibinfo {author} {\bibfnamefont {H.}~\bibnamefont {Li}}, \bibinfo {author} {\bibfnamefont {X.}~\bibnamefont {Huang}}, \bibinfo {author} {\bibfnamefont {G.}~\bibnamefont {Lu}}, \bibinfo {author} {\bibfnamefont {D.~W.}\ \bibnamefont {Fam}}, \bibinfo {author} {\bibfnamefont {A.~I.}\ \bibnamefont {Tok}}, \bibinfo {author} {\bibfnamefont {Q.}~\bibnamefont {Zhang}},\ and\ \bibinfo {author} {\bibfnamefont {H.}~\bibnamefont {Zhang}},\ }\href {https://doi.org/10.1002/smll.201101016} {\bibfield  {journal} {\bibinfo  {journal} {Small}\ }\textbf {\bibinfo {volume} {8}},\ \bibinfo {pages} {63–67} (\bibinfo {year} {2011})}\BibitemShut {NoStop}%
\bibitem [{\citenamefont {Lee}\ \emph {et~al.}(2012)\citenamefont {Lee}, \citenamefont {Zhang}, \citenamefont {Zhang}, \citenamefont {Chang}, \citenamefont {Lin}, \citenamefont {Chang}, \citenamefont {Yu}, \citenamefont {Wang}, \citenamefont {Chang}, \citenamefont {Li},\ and\ \citenamefont {Lin}}]{lee-AM-2320}%
  \BibitemOpen
  \bibfield  {author} {\bibinfo {author} {\bibfnamefont {Y.-H.}\ \bibnamefont {Lee}}, \bibinfo {author} {\bibfnamefont {X.-Q.}\ \bibnamefont {Zhang}}, \bibinfo {author} {\bibfnamefont {W.}~\bibnamefont {Zhang}}, \bibinfo {author} {\bibfnamefont {M.-T.}\ \bibnamefont {Chang}}, \bibinfo {author} {\bibfnamefont {C.-T.}\ \bibnamefont {Lin}}, \bibinfo {author} {\bibfnamefont {K.-D.}\ \bibnamefont {Chang}}, \bibinfo {author} {\bibfnamefont {Y.-C.}\ \bibnamefont {Yu}}, \bibinfo {author} {\bibfnamefont {J.~T.-W.}\ \bibnamefont {Wang}}, \bibinfo {author} {\bibfnamefont {C.-S.}\ \bibnamefont {Chang}}, \bibinfo {author} {\bibfnamefont {L.-J.}\ \bibnamefont {Li}},\ and\ \bibinfo {author} {\bibfnamefont {T.-W.}\ \bibnamefont {Lin}},\ }\href {https://doi.org/10.1002/adma.201104798} {\bibfield  {journal} {\bibinfo  {journal} {Adv{.} Mat{.}}\ }\textbf {\bibinfo {volume} {24}},\ \bibinfo {pages} {2320–2325} (\bibinfo {year} {2012})}\BibitemShut {NoStop}%
\bibitem [{\citenamefont {Liu}\ \emph {et~al.}(2012)\citenamefont {Liu}, \citenamefont {Zhang}, \citenamefont {Lee}, \citenamefont {Lin}, \citenamefont {Chang}, \citenamefont {Su}, \citenamefont {Chang}, \citenamefont {Li}, \citenamefont {Shi}, \citenamefont {Zhang}, \citenamefont {Lai},\ and\ \citenamefont {Li}}]{liu-NL-1538}%
  \BibitemOpen
  \bibfield  {author} {\bibinfo {author} {\bibfnamefont {K.-K.}\ \bibnamefont {Liu}}, \bibinfo {author} {\bibfnamefont {W.}~\bibnamefont {Zhang}}, \bibinfo {author} {\bibfnamefont {Y.-H.}\ \bibnamefont {Lee}}, \bibinfo {author} {\bibfnamefont {Y.-C.}\ \bibnamefont {Lin}}, \bibinfo {author} {\bibfnamefont {M.-T.}\ \bibnamefont {Chang}}, \bibinfo {author} {\bibfnamefont {C.-Y.}\ \bibnamefont {Su}}, \bibinfo {author} {\bibfnamefont {C.-S.}\ \bibnamefont {Chang}}, \bibinfo {author} {\bibfnamefont {H.}~\bibnamefont {Li}}, \bibinfo {author} {\bibfnamefont {Y.}~\bibnamefont {Shi}}, \bibinfo {author} {\bibfnamefont {H.}~\bibnamefont {Zhang}}, \bibinfo {author} {\bibfnamefont {C.-S.}\ \bibnamefont {Lai}},\ and\ \bibinfo {author} {\bibfnamefont {L.-J.}\ \bibnamefont {Li}},\ }\href {https://doi.org/10.1021/nl2043612} {\bibfield  {journal} {\bibinfo  {journal} {Nano Letters}\ }\textbf {\bibinfo {volume} {12}},\ \bibinfo {pages} {1538–1544} (\bibinfo {year} {2012})}\BibitemShut {NoStop}%
\bibitem [{\citenamefont {Baugher}\ \emph {et~al.}(2013)\citenamefont {Baugher}, \citenamefont {Churchill}, \citenamefont {Yang},\ and\ \citenamefont {Jarillo-Herrero}}]{baugher-NL-4212}%
  \BibitemOpen
  \bibfield  {author} {\bibinfo {author} {\bibfnamefont {B.~W.}\ \bibnamefont {Baugher}}, \bibinfo {author} {\bibfnamefont {H.~O.}\ \bibnamefont {Churchill}}, \bibinfo {author} {\bibfnamefont {Y.}~\bibnamefont {Yang}},\ and\ \bibinfo {author} {\bibfnamefont {P.}~\bibnamefont {Jarillo-Herrero}},\ }\href {https://doi.org/10.1021/nl401916s} {\bibfield  {journal} {\bibinfo  {journal} {Nano Letters}\ }\textbf {\bibinfo {volume} {13}},\ \bibinfo {pages} {4212–4216} (\bibinfo {year} {2013})}\BibitemShut {NoStop}%
\bibitem [{\citenamefont {Ahn}\ \emph {et~al.}(2017)\citenamefont {Ahn}, \citenamefont {Parkin}, \citenamefont {Naylor}, \citenamefont {Johnson},\ and\ \citenamefont {Drndić}}]{ahn-SR-7}%
  \BibitemOpen
  \bibfield  {author} {\bibinfo {author} {\bibfnamefont {J.-H.}\ \bibnamefont {Ahn}}, \bibinfo {author} {\bibfnamefont {W.~M.}\ \bibnamefont {Parkin}}, \bibinfo {author} {\bibfnamefont {C.~H.}\ \bibnamefont {Naylor}}, \bibinfo {author} {\bibfnamefont {A.~T.}\ \bibnamefont {Johnson}},\ and\ \bibinfo {author} {\bibfnamefont {M.}~\bibnamefont {Drndić}},\ }\href {https://doi.org/10.1038/s41598-017-04350-z} {\bibfield  {journal} {\bibinfo  {journal} {Scientific Reports}\ }\textbf {\bibinfo {volume} {7}},\ \bibinfo {pages} {4075} (\bibinfo {year} {2017})}\BibitemShut {NoStop}%
\bibitem [{\citenamefont {Singh}\ and\ \citenamefont {Singh}(2019)}]{singh-PRB-121201}%
  \BibitemOpen
  \bibfield  {author} {\bibinfo {author} {\bibfnamefont {A.}~\bibnamefont {Singh}}\ and\ \bibinfo {author} {\bibfnamefont {A.~K.}\ \bibnamefont {Singh}},\ }\href {https://doi.org/10.1103/PhysRevB.99.121201} {\bibfield  {journal} {\bibinfo  {journal} {Phys. Rev. B}\ }\textbf {\bibinfo {volume} {99}},\ \bibinfo {pages} {121201} (\bibinfo {year} {2019})}\BibitemShut {NoStop}%
\bibitem [{\citenamefont {Auni}\ \emph {et~al.}(2024)\citenamefont {Auni}, \citenamefont {Manopo}, \citenamefont {Sianturi}, \citenamefont {Hadju}, \citenamefont {Ivansyah}, \citenamefont {Darma},\ and\ \citenamefont {Muttaqien}}]{auni-ACSANM-28098}%
  \BibitemOpen
  \bibfield  {author} {\bibinfo {author} {\bibfnamefont {A.~K.}\ \bibnamefont {Auni}}, \bibinfo {author} {\bibfnamefont {J.}~\bibnamefont {Manopo}}, \bibinfo {author} {\bibfnamefont {I.~S.}\ \bibnamefont {Sianturi}}, \bibinfo {author} {\bibfnamefont {A.}~\bibnamefont {Hadju}}, \bibinfo {author} {\bibfnamefont {A.~L.}\ \bibnamefont {Ivansyah}}, \bibinfo {author} {\bibfnamefont {Y.}~\bibnamefont {Darma}},\ and\ \bibinfo {author} {\bibfnamefont {F.}~\bibnamefont {Muttaqien}},\ }\href {https://doi.org/10.1021/acsanm.4c04688} {\bibfield  {journal} {\bibinfo  {journal} {ACS Appl{.} Nano Mat{.}}\ }\textbf {\bibinfo {volume} {7}},\ \bibinfo {pages} {28098–28105} (\bibinfo {year} {2024})}\BibitemShut {NoStop}%
\bibitem [{\citenamefont {{He}}\ \emph {et~al.}(2010)\citenamefont {{He}}, \citenamefont {{Wu}}, \citenamefont {{Sa}}, \citenamefont {{Li}},\ and\ \citenamefont {{Wei}}}]{he-APL-96}%
  \BibitemOpen
  \bibfield  {author} {\bibinfo {author} {\bibfnamefont {J.}~\bibnamefont {{He}}}, \bibinfo {author} {\bibfnamefont {K.}~\bibnamefont {{Wu}}}, \bibinfo {author} {\bibfnamefont {R.}~\bibnamefont {{Sa}}}, \bibinfo {author} {\bibfnamefont {Q.}~\bibnamefont {{Li}}},\ and\ \bibinfo {author} {\bibfnamefont {Y.}~\bibnamefont {{Wei}}},\ }\bibfield  {journal} {\bibinfo  {journal} {Appl{.} Phys{.} Lett{.}}\ }\textbf {\bibinfo {volume} {96}},\ \href {https://doi.org/10.1063/1.3318254} {10.1063/1.3318254} (\bibinfo {year} {2010})\BibitemShut {NoStop}%
\bibitem [{\citenamefont {{Li}}\ \emph {et~al.}(2015)\citenamefont {{Li}}, \citenamefont {{Fang}}, \citenamefont {{Wu}},\ and\ \citenamefont {{Zhu}}}]{li-AIPA-5}%
  \BibitemOpen
  \bibfield  {author} {\bibinfo {author} {\bibfnamefont {X.~D.}\ \bibnamefont {{Li}}}, \bibinfo {author} {\bibfnamefont {Y.~M.}\ \bibnamefont {{Fang}}}, \bibinfo {author} {\bibfnamefont {S.~Q.}\ \bibnamefont {{Wu}}},\ and\ \bibinfo {author} {\bibfnamefont {Z.~Z.}\ \bibnamefont {{Zhu}}},\ }\href {https://doi.org/10.1063/1.4921564} {\bibfield  {journal} {\bibinfo  {journal} {AIP Adv{.}}\ }\textbf {\bibinfo {volume} {5}},\ \bibinfo {pages} {082504} (\bibinfo {year} {2015})}\BibitemShut {NoStop}%
\bibitem [{\citenamefont {Aghajanian}\ \emph {et~al.}(2018)\citenamefont {Aghajanian}, \citenamefont {Mostofi},\ and\ \citenamefont {Lischner}}]{aghajanian-ARXIV-1805}%
  \BibitemOpen
  \bibfield  {author} {\bibinfo {author} {\bibfnamefont {M.}~\bibnamefont {Aghajanian}}, \bibinfo {author} {\bibfnamefont {A.~A.}\ \bibnamefont {Mostofi}},\ and\ \bibinfo {author} {\bibfnamefont {J.}~\bibnamefont {Lischner}},\ }\href {https://doi.org/10.48550/arXiv.1805.02167} {\bibinfo {title} {Bound states of charged adatoms on {\mos}: Screening and multivalley effects}} (\bibinfo {year} {2018}),\ \Eprint {https://arxiv.org/abs/1805.02167} {arXiv:1805.02167 [cond-mat.mes-hall]} \BibitemShut {NoStop}%
\bibitem [{\citenamefont {{Ataca}}\ and\ \citenamefont {{Ciraci}}(2011)}]{ataca-JPCC-115-13303}%
  \BibitemOpen
  \bibfield  {author} {\bibinfo {author} {\bibfnamefont {C.}~\bibnamefont {{Ataca}}}\ and\ \bibinfo {author} {\bibfnamefont {S.}~\bibnamefont {{Ciraci}}},\ }\href {https://doi.org/10.1021/jp2000442} {\bibfield  {journal} {\bibinfo  {journal} {J{.} Phys{.} Chem{.} C}\ }\textbf {\bibinfo {volume} {115}},\ \bibinfo {pages} {13303–13311} (\bibinfo {year} {2011})}\BibitemShut {NoStop}%
\bibitem [{\citenamefont {Rayson}\ and\ \citenamefont {Briddon}(2009)}]{rayson-PRB-80-205104}%
  \BibitemOpen
  \bibfield  {author} {\bibinfo {author} {\bibfnamefont {M.~J.}\ \bibnamefont {Rayson}}\ and\ \bibinfo {author} {\bibfnamefont {P.~R.}\ \bibnamefont {Briddon}},\ }\href@noop {} {\bibfield  {journal} {\bibinfo  {journal} {Phys{.} Rev{.} B}\ }\textbf {\bibinfo {volume} {80}},\ \bibinfo {pages} {205104} (\bibinfo {year} {2009})}\BibitemShut {NoStop}%
\bibitem [{\citenamefont {Jones}\ and\ \citenamefont {Briddon}(1998)}]{jones-SSM-51-287}%
  \BibitemOpen
  \bibfield  {author} {\bibinfo {author} {\bibfnamefont {R.}~\bibnamefont {Jones}}\ and\ \bibinfo {author} {\bibfnamefont {P.~R.}\ \bibnamefont {Briddon}},\ }\href@noop {} {\bibfield  {journal} {\bibinfo  {journal} {Semiconductors and Semimetals}\ }\textbf {\bibinfo {volume} {51}},\ \bibinfo {pages} {287} (\bibinfo {year} {1998})}\BibitemShut {NoStop}%
\bibitem [{\citenamefont {Perdew}\ \emph {et~al.}(1996)\citenamefont {Perdew}, \citenamefont {Burke},\ and\ \citenamefont {Ernzerhof}}]{perdew-PRL-77-3865}%
  \BibitemOpen
  \bibfield  {author} {\bibinfo {author} {\bibfnamefont {J.~P.}\ \bibnamefont {Perdew}}, \bibinfo {author} {\bibfnamefont {K.}~\bibnamefont {Burke}},\ and\ \bibinfo {author} {\bibfnamefont {M.}~\bibnamefont {Ernzerhof}},\ }\href@noop {} {\bibfield  {journal} {\bibinfo  {journal} {Phys{.} Rev{.} Lett{.}}\ }\textbf {\bibinfo {volume} {77}},\ \bibinfo {pages} {3865} (\bibinfo {year} {1996})}\BibitemShut {NoStop}%
\bibitem [{\citenamefont {Grimme}(2006)}]{grimme-JCC-1787}%
  \BibitemOpen
  \bibfield  {author} {\bibinfo {author} {\bibfnamefont {S.}~\bibnamefont {Grimme}},\ }\href {https://doi.org/10.1002/jcc.20495} {\bibfield  {journal} {\bibinfo  {journal} {J{.} Comp{.} Chem{.}}\ }\textbf {\bibinfo {volume} {27}},\ \bibinfo {pages} {1787} (\bibinfo {year} {2006})}\BibitemShut {NoStop}%
\bibitem [{\citenamefont {Hartwigsen}\ \emph {et~al.}(1998)\citenamefont {Hartwigsen}, \citenamefont {Goedecker},\ and\ \citenamefont {Hutter}}]{hartwigsen-PRB-58-3641}%
  \BibitemOpen
  \bibfield  {author} {\bibinfo {author} {\bibfnamefont {C.}~\bibnamefont {Hartwigsen}}, \bibinfo {author} {\bibfnamefont {S.}~\bibnamefont {Goedecker}},\ and\ \bibinfo {author} {\bibfnamefont {J.}~\bibnamefont {Hutter}},\ }\href@noop {} {\bibfield  {journal} {\bibinfo  {journal} {Phys{.} Rev{.} B}\ }\textbf {\bibinfo {volume} {58}},\ \bibinfo {pages} {3641} (\bibinfo {year} {1998})}\BibitemShut {NoStop}%
\bibitem [{\citenamefont {Krack}(2005)}]{krack-TCA-114-145}%
  \BibitemOpen
  \bibfield  {author} {\bibinfo {author} {\bibfnamefont {M.}~\bibnamefont {Krack}},\ }\href {https://doi.org/10.1007/s00214-005-0655-y} {\bibfield  {journal} {\bibinfo  {journal} {Theor{.} Chem{.} Acc{.}}\ }\textbf {\bibinfo {volume} {114}},\ \bibinfo {pages} {145} (\bibinfo {year} {2005})}\BibitemShut {NoStop}%
\bibitem [{\citenamefont {Goss}\ \emph {et~al.}(2007)\citenamefont {Goss}, \citenamefont {Shaw},\ and\ \citenamefont {Briddon}}]{goss-TAP-104-69}%
  \BibitemOpen
  \bibfield  {author} {\bibinfo {author} {\bibfnamefont {J.~P.}\ \bibnamefont {Goss}}, \bibinfo {author} {\bibfnamefont {M.~J.}\ \bibnamefont {Shaw}},\ and\ \bibinfo {author} {\bibfnamefont {P.~R.}\ \bibnamefont {Briddon}},\ }in\ \href@noop {} {\emph {\bibinfo {booktitle} {Theory of Defects in Semiconductors}}},\ \bibinfo {series} {Topics in Applied Physics}, Vol.\ \bibinfo {volume} {104},\ \bibinfo {editor} {edited by\ \bibinfo {editor} {\bibfnamefont {D.~A.}\ \bibnamefont {Drabold}}\ and\ \bibinfo {editor} {\bibfnamefont {S.~K.}\ \bibnamefont {Estreicher}}}\ (\bibinfo {year} {2007})\ pp.\ \bibinfo {pages} {69--94}\BibitemShut {NoStop}%
\bibitem [{\citenamefont {Monkhorst}\ and\ \citenamefont {Pack}(1976)}]{monkhorst-PRB-13-5188}%
  \BibitemOpen
  \bibfield  {author} {\bibinfo {author} {\bibfnamefont {H.~J.}\ \bibnamefont {Monkhorst}}\ and\ \bibinfo {author} {\bibfnamefont {J.~D.}\ \bibnamefont {Pack}},\ }\href@noop {} {\bibfield  {journal} {\bibinfo  {journal} {Phys{.} Rev{.} B}\ }\textbf {\bibinfo {volume} {13}},\ \bibinfo {pages} {5188} (\bibinfo {year} {1976})}\BibitemShut {NoStop}%
\bibitem [{\citenamefont {Le}\ \emph {et~al.}(2020)\citenamefont {Le}, \citenamefont {Chihaia}, \citenamefont {Pham-Ho},\ and\ \citenamefont {Son}}]{le-RSCA-2046}%
  \BibitemOpen
  \bibfield  {author} {\bibinfo {author} {\bibfnamefont {O.~K.}\ \bibnamefont {Le}}, \bibinfo {author} {\bibfnamefont {V.}~\bibnamefont {Chihaia}}, \bibinfo {author} {\bibfnamefont {M.-P.}\ \bibnamefont {Pham-Ho}},\ and\ \bibinfo {author} {\bibfnamefont {D.~N.}\ \bibnamefont {Son}},\ }\href {https://doi.org/10.1039/C9RA09042H} {\bibfield  {journal} {\bibinfo  {journal} {Royal Soc{.} Chem{.} Adv{.}}\ }\textbf {\bibinfo {volume} {10}},\ \bibinfo {pages} {4201} (\bibinfo {year} {2020})}\BibitemShut {NoStop}%
\bibitem [{\citenamefont {Ataca}\ \emph {et~al.}(2011)\citenamefont {Ataca}, \citenamefont {Sahin}, \citenamefont {Akturk},\ and\ \citenamefont {Ciraci}}]{ataca2011mechanical}%
  \BibitemOpen
  \bibfield  {author} {\bibinfo {author} {\bibfnamefont {C.}~\bibnamefont {Ataca}}, \bibinfo {author} {\bibfnamefont {H.}~\bibnamefont {Sahin}}, \bibinfo {author} {\bibfnamefont {E.}~\bibnamefont {Akturk}},\ and\ \bibinfo {author} {\bibfnamefont {S.}~\bibnamefont {Ciraci}},\ }\href@noop {} {\bibfield  {journal} {\bibinfo  {journal} {J{.} Phys{.} Chem{.} C}\ }\textbf {\bibinfo {volume} {115}},\ \bibinfo {pages} {3934} (\bibinfo {year} {2011})}\BibitemShut {NoStop}%
\bibitem [{\citenamefont {Hieu}\ \emph {et~al.}(2018)\citenamefont {Hieu}, \citenamefont {Ilyasov}, \citenamefont {Vu}, \citenamefont {Poklonski}, \citenamefont {Phuc}, \citenamefont {Phuong}, \citenamefont {Hoi},\ and\ \citenamefont {Nguyen}}]{hieu2018first}%
  \BibitemOpen
  \bibfield  {author} {\bibinfo {author} {\bibfnamefont {N.~N.}\ \bibnamefont {Hieu}}, \bibinfo {author} {\bibfnamefont {V.~V.}\ \bibnamefont {Ilyasov}}, \bibinfo {author} {\bibfnamefont {T.~V.}\ \bibnamefont {Vu}}, \bibinfo {author} {\bibfnamefont {N.~A.}\ \bibnamefont {Poklonski}}, \bibinfo {author} {\bibfnamefont {H.~V.}\ \bibnamefont {Phuc}}, \bibinfo {author} {\bibfnamefont {L.~T.}\ \bibnamefont {Phuong}}, \bibinfo {author} {\bibfnamefont {B.~D.}\ \bibnamefont {Hoi}},\ and\ \bibinfo {author} {\bibfnamefont {C.~V.}\ \bibnamefont {Nguyen}},\ }\href {https://doi.org/10.1016/j.spmi.2018.01.012} {\bibfield  {journal} {\bibinfo  {journal} {Superlattices and Microstructures}\ }\textbf {\bibinfo {volume} {115}},\ \bibinfo {pages} {10} (\bibinfo {year} {2018})}\BibitemShut {NoStop}%
\bibitem [{\citenamefont {Zhang}\ and\ \citenamefont {Northrup}(1991)}]{zhang-PRL-67-2339}%
  \BibitemOpen
  \bibfield  {author} {\bibinfo {author} {\bibfnamefont {S.~B.}\ \bibnamefont {Zhang}}\ and\ \bibinfo {author} {\bibfnamefont {J.~E.}\ \bibnamefont {Northrup}},\ }\href@noop {} {\bibfield  {journal} {\bibinfo  {journal} {Phys{.} Rev{.} Lett{.}}\ }\textbf {\bibinfo {volume} {67}},\ \bibinfo {pages} {2339} (\bibinfo {year} {1991})}\BibitemShut {NoStop}%
\bibitem [{\citenamefont {Goto}(1966)}]{goto-JPSJ-895}%
  \BibitemOpen
  \bibfield  {author} {\bibinfo {author} {\bibfnamefont {F.}~\bibnamefont {Goto}},\ }\href {https://doi.org/10.1143/jpsj.21.895} {\bibfield  {journal} {\bibinfo  {journal} {J{.} Phys{.} Soc{.} Jap{.}}\ }\textbf {\bibinfo {volume} {21}},\ \bibinfo {pages} {895–906} (\bibinfo {year} {1966})}\BibitemShut {NoStop}%
\bibitem [{\citenamefont {Gaydon}\ and\ \citenamefont {Penney}(1945)}]{gaydon-PRSLA-374}%
  \BibitemOpen
  \bibfield  {author} {\bibinfo {author} {\bibfnamefont {A.~G.}\ \bibnamefont {Gaydon}}\ and\ \bibinfo {author} {\bibfnamefont {W.~G.}\ \bibnamefont {Penney}},\ }\href {https://doi.org/10.1098/rspa.1945.0009} {\bibfield  {journal} {\bibinfo  {journal} {Proc{.} R{.} Soc{.} London, Ser{.} A}\ }\textbf {\bibinfo {volume} {183}},\ \bibinfo {pages} {374–388} (\bibinfo {year} {1945})}\BibitemShut {NoStop}%
\bibitem [{\citenamefont {Henkelman}\ \emph {et~al.}(2000)\citenamefont {Henkelman}, \citenamefont {Uberuaga},\ and\ \citenamefont {J{\'o}nsson}}]{henkelman-JCP-113-9901}%
  \BibitemOpen
  \bibfield  {author} {\bibinfo {author} {\bibfnamefont {G.}~\bibnamefont {Henkelman}}, \bibinfo {author} {\bibfnamefont {B.~P.}\ \bibnamefont {Uberuaga}},\ and\ \bibinfo {author} {\bibfnamefont {H.}~\bibnamefont {J{\'o}nsson}},\ }\href@noop {} {\bibfield  {journal} {\bibinfo  {journal} {J{.} Chem{.} Phys{.}}\ }\textbf {\bibinfo {volume} {113}},\ \bibinfo {pages} {9901} (\bibinfo {year} {2000})}\BibitemShut {NoStop}%
\bibitem [{\citenamefont {Henkelman}\ and\ \citenamefont {J{\'o}nsson}(2000)}]{henkelman-JCP-113-9978}%
  \BibitemOpen
  \bibfield  {author} {\bibinfo {author} {\bibfnamefont {G.}~\bibnamefont {Henkelman}}\ and\ \bibinfo {author} {\bibfnamefont {H.}~\bibnamefont {J{\'o}nsson}},\ }\href@noop {} {\bibfield  {journal} {\bibinfo  {journal} {J{.} Chem{.} Phys{.}}\ }\textbf {\bibinfo {volume} {113}},\ \bibinfo {pages} {9978} (\bibinfo {year} {2000})}\BibitemShut {NoStop}%
\bibitem [{\citenamefont {Jones}\ \emph {et~al.}(1994)\citenamefont {Jones}, \citenamefont {Goss}, \citenamefont {Ewels},\ and\ \citenamefont {{\"O}berg}}]{jones-PRB-50-8378}%
  \BibitemOpen
  \bibfield  {author} {\bibinfo {author} {\bibfnamefont {R.}~\bibnamefont {Jones}}, \bibinfo {author} {\bibfnamefont {J.}~\bibnamefont {Goss}}, \bibinfo {author} {\bibfnamefont {C.}~\bibnamefont {Ewels}},\ and\ \bibinfo {author} {\bibfnamefont {S.}~\bibnamefont {{\"O}berg}},\ }\href@noop {} {\bibfield  {journal} {\bibinfo  {journal} {Phys{.} Rev{.} B}\ }\textbf {\bibinfo {volume} {50}},\ \bibinfo {pages} {8378} (\bibinfo {year} {1994})}\BibitemShut {NoStop}%
\bibitem [{\citenamefont {KuO}\ and\ \citenamefont {H{\"a}gg}(1952)}]{kuo-N-170-245}%
  \BibitemOpen
  \bibfield  {author} {\bibinfo {author} {\bibfnamefont {K.}~\bibnamefont {KuO}}\ and\ \bibinfo {author} {\bibfnamefont {G.}~\bibnamefont {H{\"a}gg}},\ }\href {https://doi.org/10.1038/170245a0} {\bibfield  {journal} {\bibinfo  {journal} {Nature}\ }\textbf {\bibinfo {volume} {170}},\ \bibinfo {pages} {245} (\bibinfo {year} {1952})}\BibitemShut {NoStop}%
\bibitem [{\citenamefont {{Cho}}\ \emph {et~al.}(2023)\citenamefont {{Cho}}, \citenamefont {{Sim}}, \citenamefont {{Lee}}, \citenamefont {{Hoang}},\ and\ \citenamefont {{Seong}}}]{Cho_Sim_Lee_Hoang_Seong_2023}%
  \BibitemOpen
  \bibfield  {author} {\bibinfo {author} {\bibfnamefont {Y.~J.}\ \bibnamefont {{Cho}}}, \bibinfo {author} {\bibfnamefont {Y.}~\bibnamefont {{Sim}}}, \bibinfo {author} {\bibfnamefont {J.-H.}\ \bibnamefont {{Lee}}}, \bibinfo {author} {\bibfnamefont {N.~T.}\ \bibnamefont {{Hoang}}},\ and\ \bibinfo {author} {\bibfnamefont {M.-J.}\ \bibnamefont {{Seong}}},\ }\href {https://doi.org/10.1016/j.cap.2022.11.008} {\bibfield  {journal} {\bibinfo  {journal} {Current Applied Physics}\ }\textbf {\bibinfo {volume} {45}},\ \bibinfo {pages} {99–104} (\bibinfo {year} {2023})}\BibitemShut {NoStop}%
\bibitem [{\citenamefont {{Zhu}}\ \emph {et~al.}(2017)\citenamefont {{Zhu}}, \citenamefont {{Shu}}, \citenamefont {{Jiang}}, \citenamefont {{Lv}}, \citenamefont {{Asokan}}, \citenamefont {{Omar}}, \citenamefont {{Yuan}}, \citenamefont {{Zhang}},\ and\ \citenamefont {{Jin}}}]{Zhu_Shu_Jiang_Lv_Asokan_Omar_Yuan_Zhang_Jin_2017}%
  \BibitemOpen
  \bibfield  {author} {\bibinfo {author} {\bibfnamefont {D.}~\bibnamefont {{Zhu}}}, \bibinfo {author} {\bibfnamefont {H.}~\bibnamefont {{Shu}}}, \bibinfo {author} {\bibfnamefont {F.}~\bibnamefont {{Jiang}}}, \bibinfo {author} {\bibfnamefont {D.}~\bibnamefont {{Lv}}}, \bibinfo {author} {\bibfnamefont {V.}~\bibnamefont {{Asokan}}}, \bibinfo {author} {\bibfnamefont {O.}~\bibnamefont {{Omar}}}, \bibinfo {author} {\bibfnamefont {J.}~\bibnamefont {{Yuan}}}, \bibinfo {author} {\bibfnamefont {Z.}~\bibnamefont {{Zhang}}},\ and\ \bibinfo {author} {\bibfnamefont {C.}~\bibnamefont {{Jin}}},\ }\href {https://doi.org/10.1038/s41699-017-0010-x} {\bibfield  {journal} {\bibinfo  {journal} {npj 2D Mat{.} and Appl{.}}\ }\textbf {\bibinfo {volume} {1}},\ \bibinfo {pages} {8} (\bibinfo {year} {2017})}\BibitemShut {NoStop}%
\bibitem [{\citenamefont {{Yue}}\ \emph {et~al.}(2019)\citenamefont {{Yue}}, \citenamefont {{Jian}}, \citenamefont {{Dong}}, \citenamefont {{Luo}},\ and\ \citenamefont {{Chang}}}]{Yue_Jian_Dong_Luo_Chang_2019}%
  \BibitemOpen
  \bibfield  {author} {\bibinfo {author} {\bibfnamefont {J.}~\bibnamefont {{Yue}}}, \bibinfo {author} {\bibfnamefont {J.}~\bibnamefont {{Jian}}}, \bibinfo {author} {\bibfnamefont {P.}~\bibnamefont {{Dong}}}, \bibinfo {author} {\bibfnamefont {L.}~\bibnamefont {{Luo}}},\ and\ \bibinfo {author} {\bibfnamefont {F.}~\bibnamefont {{Chang}}},\ }\href {https://doi.org/10.1088/1757-899x/592/1/012044} {\bibfield  {journal} {\bibinfo  {journal} {IOP Mater{.} Sci{.} Eng{.}}\ }\textbf {\bibinfo {volume} {592}},\ \bibinfo {pages} {012044} (\bibinfo {year} {2019})}\BibitemShut {NoStop}%
\bibitem [{\citenamefont {Brewer}(1975)}]{brewer-BERK-1975}%
  \BibitemOpen
  \bibfield  {author} {\bibinfo {author} {\bibfnamefont {L.}~\bibnamefont {Brewer}},\ }\href {https://doi.org/10.2172/7187973} {\emph {\bibinfo {title} {Cohesive energies of the elements}}},\ \bibinfo {type} {Tech. Rep.}\ (\bibinfo  {institution} {Lawrence Berkeley National Lab. (LBNL), Berkeley, CA (United States)},\ \bibinfo {year} {1975})\BibitemShut {NoStop}%
\bibitem [{\citenamefont {Feng}\ \emph {et~al.}(2014)\citenamefont {Feng}, \citenamefont {Su},\ and\ \citenamefont {Liu}}]{feng-JAC-122}%
  \BibitemOpen
  \bibfield  {author} {\bibinfo {author} {\bibfnamefont {L.-p.}\ \bibnamefont {Feng}}, \bibinfo {author} {\bibfnamefont {J.}~\bibnamefont {Su}},\ and\ \bibinfo {author} {\bibfnamefont {Z.-t.}\ \bibnamefont {Liu}},\ }\href {https://doi.org/10.1016/j.jallcom.2014.06.018} {\bibfield  {journal} {\bibinfo  {journal} {J{.} Appl{.} Cryst{.}}\ }\textbf {\bibinfo {volume} {613}},\ \bibinfo {pages} {122} (\bibinfo {year} {2014})}\BibitemShut {NoStop}%
\bibitem [{\citenamefont {Feng}\ \emph {et~al.}(2018)\citenamefont {Feng}, \citenamefont {Sun}, \citenamefont {Li}, \citenamefont {Su}, \citenamefont {Zhang},\ and\ \citenamefont {Liu}}]{feng-MCP-146}%
  \BibitemOpen
  \bibfield  {author} {\bibinfo {author} {\bibfnamefont {L.-p.}\ \bibnamefont {Feng}}, \bibinfo {author} {\bibfnamefont {H.-q.}\ \bibnamefont {Sun}}, \bibinfo {author} {\bibfnamefont {A.}~\bibnamefont {Li}}, \bibinfo {author} {\bibfnamefont {J.}~\bibnamefont {Su}}, \bibinfo {author} {\bibfnamefont {Y.}~\bibnamefont {Zhang}},\ and\ \bibinfo {author} {\bibfnamefont {Z.-t.}\ \bibnamefont {Liu}},\ }\href {https://doi.org/10.1016/j.matchemphys.2018.01.015} {\bibfield  {journal} {\bibinfo  {journal} {Mater{.} Chem{.} Phys{.}}\ }\textbf {\bibinfo {volume} {209}},\ \bibinfo {pages} {146–151} (\bibinfo {year} {2018})}\BibitemShut {NoStop}%
\bibitem [{\citenamefont {{Wolff}}\ and\ \citenamefont {{Szydlowski}}(1985)}]{wolff-CJC-63-1708}%
  \BibitemOpen
  \bibfield  {author} {\bibinfo {author} {\bibfnamefont {H.}~\bibnamefont {{Wolff}}}\ and\ \bibinfo {author} {\bibfnamefont {J.}~\bibnamefont {{Szydlowski}}},\ }\href {https://doi.org/10.1139/v85-287} {\bibfield  {journal} {\bibinfo  {journal} {Canadian J{.} Chem{.}}\ }\textbf {\bibinfo {volume} {63}},\ \bibinfo {pages} {1708–1712} (\bibinfo {year} {1985})}\BibitemShut {NoStop}%
\end{thebibliography}%

\end{document}